\documentclass[a4paper,10pt]{article}
\usepackage[margin=1in]{geometry}
\usepackage{amsmath,amsfonts,epsfig,amssymb, amsthm}
\usepackage{epsfig,graphicx,color,longtable}
\usepackage[T1]{fontenc}
 \usepackage{graphicx}
\usepackage{appendix}
\usepackage{mathtools}
\usepackage{longtable}
\usepackage[ruled,vlined]{algorithm2e}
\usepackage{enumitem}
\usepackage{lscape}
\usepackage{caption}
\usepackage{blindtext}
\usepackage{color,soul}
\title{Optimal Strokes of Low Reynolds Number Linked-Sphere Swimmers}
\author{Qixuan Wang  \\
	Department of Mathematics\\
	University of California at Riverside\\
	Riverside, CA 925521, USA\\
	qixuanw@ucr.edu
	}

\date{\today}

\begin{document}

\maketitle

\begin{abstract}
Optimal gait design is important for micro-organisms and micro-robots that propel themselves in a
fluid environment in the absence of external force or torque. The simplest models of shape changes are those
that comprise a series of linked-spheres that can change their separation and/or their sizes.  
We examine the dynamics of three existing linked-sphere types of modeling swimmers in low Reynolds number 
Newtonian fluids using asymptotic analysis, and obatain
their optimal swimming strokes by solving the Euler--Lagrange
equation using the shooting method. The~numerical results reveal that (1) with the minimal 2 degrees of freedom in
shape deformations, the~model swimmer adopting the mixed shape deformation modes strategy is more efficient than those
with a single-mode of shape deformation modes, and (2) the swimming efficiency mostly decreases as the number of spheres
increases, indicating that more degrees of freedom in shape deformations might not be a good strategy in optimal gait 
design in low Reynolds number locomotion.
\end{abstract}

\section{Introduction}

Swimming by shape changes at low Reynolds number (LRN) is widely used in biology and micro-robotic design.
In this flow regime, inertial effects are negligible, and the micro-organisms or micro-robots propel themselves by exploiting the
viscous resistance of the fluid. For example, while~a scallop that can only open or close its shell can swim in the ocean 
by accelerating the surrounding water, such a swimming strategy does not work at LRN, which 
is generalized by the principle: any~time-reversible swimming stroke leads to no net translation at LRN Newtonian 
fluid, known as the \textit{scallop theorem}~\cite{purcell1977life}.

It is important to understand how the performance of swimming depends on the geometric patterns of shape 
deformations for micro-swimmers. { In microbiology,} to fight the viscous resistance, different microorganisms adopt various
propulsion mechanisms and directed locomotion strategies for searching for food and running from predators.
For example, individual cells such as bacteria find food by a combination of taxis and kinesis using a flagellated 
or ciliated mode of swimming~\cite{sokolov2010swimming,riedel2005self,lauga2006swimming,rafai2010effective}. 
Recently,~it~was discovered that Dd cells can occasionally detach from the substrate 
and stay completely free in suspension for a few minutes before they slowly sink; during the free suspension stage, 
cells~continue to form pseudopods that convert to rear-ward moving bumps, thereby~propelling the cell through the 
surrounding fluid in a totally adhesion- free fashion~\cite{van2011amoeboid}. Human neutrophils can swim to a chemoattractant fMLP 
(formyl-methionylleucyl-phenylalanine) source at a speed similar to that of cells migrating on a glass coverslip under similar conditions
\cite{barry2010dictyostelium}. Most recently and equally striking, Drosophila fat body cells 
can actively swim to wounds in an adhesion- independent motility mode associated with actomyosin-driven, peristaltic 
cell shape deformations~\cite{franz2018fat}. {In~micro bio-engineering, medical microrobots revolutionize
many aspects of medicine in recent years, which~make existing therapeutic and diagnostic procedures less invasive
\cite{nelson2010microrobots}.}
Different LRN swimming models and micro-robots have been designed since
Purcell's two-hinge model was advanced~\cite{purcell1977life}. In particular, various linked-sphere types of models
have appeared, since their simple geometry permits both analytical and computational results~\cite{najafi2004simple,golestanian2008analytic, 
alexander2009hydrodynamics,avron2005pushmepullyou,alouges2018parking, rizvi2018three,wang2012models,curtis2013three}. 
{ These analytical and numerical results have greatly inspired the designs of micro robotic devices, for example,
swimmers with 2 and 3 rotatory cylinders have been built to study the hydrodynamic interaction between 
a wall and an active swimmer~\cite{zhang2010experimental,or2011dynamics}. Other micro-robots inspired
by analytical/numerical works include the Quadroar swimmer, which~consists of rotating disks and a linear
actuator~\cite{saadat2019experimental}, as well as the Purcell's two-hinge model~\cite{kadam2017trajectory}.
}

An important problem in LRN swimming is to find the optimized swimming stroke of the micro-swimmer, either 
(1) with respect to time, i.e., the stroke in one swimming cycle that moves the cell farthest, or (2) with respect to energy, 
i.e., among all strokes with designated starting and end points, find the one that consumes the least energy. 
These are usually called the \textit{time optimal control} and the \textit{energy optimal control} problems, respectively. 
Both optimal problems have attracted substantial interest in optimization as well as geometry~\cite{lighthill1952squirming,blake1971self,shapere1989efficiencies,
alouges2008optimal, alouges2009optimal,avron2004optimal,avron2008geometric,ishimoto2014swimming,alouges2011numerical,loheac2013controllability, 
loheac2014controllability, chambrion2019optimal}. Moreover, recently techniques from machine learning have been
applied to linked-sphere types of model gait design, which allows incorporating environmental influences on the micro-swimmer's
swimming behavior, including noise and a frictional medium~\cite{tsang2018self}.

\begin{figure}[h]
\centering
\includegraphics[width=0.9\textwidth]{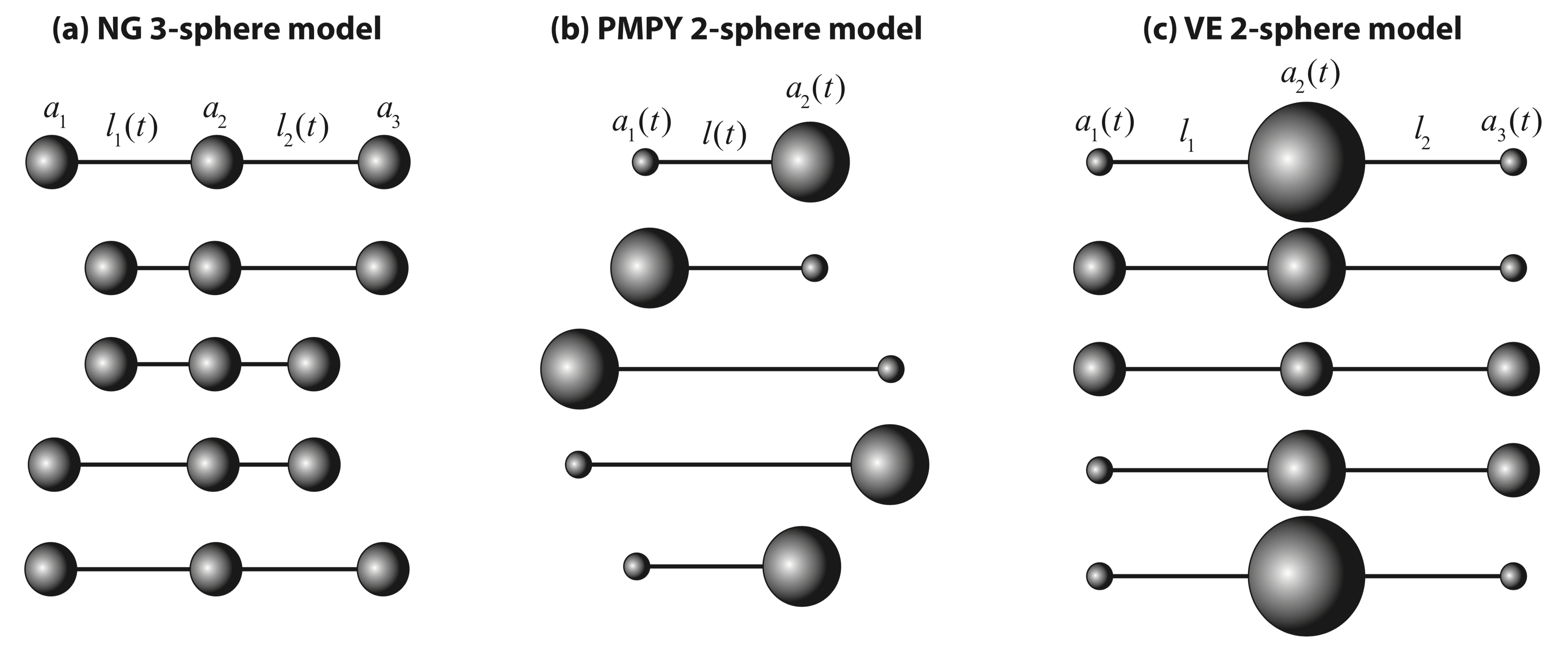}
\caption{\textbf{{Linked-sphere LRN swimmers}.} (\textbf{a}) a NG 3-sphere model~\cite{najafi2004simple, golestanian2008analytic, alexander2009hydrodynamics}. 
(\textbf{b}) a PMPY 2-sphere model~\cite{avron2005pushmepullyou}. (\textbf{{c}}) 
a
 VE 3-sphere model~\cite{wang2012models}. Figures show example strokes of each
swimmer. Only shape deformations are presented, net translation of the swimmer is not shown.}
\label{Fig.NG-PMPY-VE}
\end{figure}

Here by investigating the optimal strokes of a group of linked-sphere types of LRN modeling swimmers,
we study the efficiencies of propelling mechanisms at LRN of different types of shape deformations. 
We start from three linked-sphere swimmers--Najafi-Golestanian (NG) 3-sphere accordion model (Figure \ref{Fig.NG-PMPY-VE}a)
\cite{najafi2004simple, golestanian2008analytic, alexander2009hydrodynamics}, \textit{pushmepullyou} (PMPY) 2-sphere model (Figure \ref{Fig.NG-PMPY-VE}b)
\cite{avron2005pushmepullyou}, and~the volume-exchange (VE) 3-sphere model (Figure \ref{Fig.NG-PMPY-VE}c)~\cite{wang2012models}. 
All three models have only 2 degrees of freedom in their shape deformations, which is a minimal requirement
that enables the swimmer to propel itself at LRN, according to the \textit{scallop theorem}~\cite{purcell1977life}. 
In particular, the shape deformations in the three models can be generalized as body elongation and/or mass
transportation, and we investigate the efficiencies of these two shape deformation modes on the swimming 
performances. Then we generalize our results into a linear chain of spheres. The results are presented as follows:
in Section~\ref{Sec.IntroSwim} we present a brief introduction of the LRN swimming problem; in Section \ref{Sec.LinkSphBasic}
we review the three existing linked-sphere types of swimmers: NG 3-sphere, PMPY 2-sphere and VE 3-sphere models
and discuss their optimization problems in Section \ref{Sec.OptBasic}; finally in Section \ref{Sec.OptChain} we discuss the optimization problem
of models consisting of a chain of spheres.

\section{Swimming at LRN by Shape Changes-the Exterior Problem\label{Sec.IntroSwim}}
The Navier-Stokes equations for an incompressible fluid of density $\rho$, viscosity $\mu$, and velocity $\mathbf{u}$~are
\begin{eqnarray*}
\rho \dfrac{\partial \mathbf{u}}{\partial t} + \rho \big( \mathbf{u} \cdot \nabla  \big) \mathbf{u} &=& - \nabla p + \mu \Delta \mathbf{u} + \mathbf{f}_{\textrm{ext}} \\
\nabla \cdot \mathbf{u} &=& 0
\end{eqnarray*}
where $\mathbf{f}_{\textrm{ext}} $ is the external force field. Herein we assume that the swimmer is self-propelled and does not 
rely on any exterior force, and therefore we require that $\mathbf{f}_{\textrm{ext}}  = \mathbf{0}$. The Reynolds number based on
a characteristic length scale $L$ and speed scale $U$ is $\textrm{Re} = \rho LU/\mu$, and when converted to dimensionless form
and the symbols re-defined, the equations read
\begin{eqnarray}\label{Eq.ReSl-NS}
\textrm{Re} \Big[ \textrm{Sl} \dfrac{\partial \mathbf{u}}{\partial t} + \big( \mathbf{u} \cdot \nabla \big) \mathbf{u} \Big] &=& - \nabla p + \Delta \mathbf{u} \\ \nonumber
\nabla \cdot \mathbf{u} &=& 0 
\end{eqnarray}

Here $\textrm{Sl} = \omega L / U$ is the Strouhal number and $\omega$ is a characteristic frequency of the shape changes.
When $\textrm{Re} \ll 1$ the convective momentum term in Equation (\ref{Eq.ReSl-NS}) can be neglected, but the time variation
requires that $\textrm{ReSl} = \omega L^2 / \nu \ll 1$. When both terms are neglected, which we assume throughout, the~flow
is governed by the Stokes equations:
\begin{eqnarray}\label{Eq.Stokes}
\mu \Delta \mathbf{u} - \nabla p = \mathbf{0}, \qquad \nabla \cdot \mathbf{u} = 0
\end{eqnarray}

We also consider the propulsion problem in an infinite domain and impose the condition $\mathbf{u} |_{\mathbf{x} \rightarrow \infty} = \mathbf{0}$
on the velocity field.
 
In the LRN regime time does not appear explicitly, momentum is assumed to equilibrate instantaneously, and bodies 
move by exploiting the viscous resistance of the fluid. As a result, time-reversible deformations produce no motion, 
which is the content of the \textit{scallop theorem}~\cite{purcell1977life}. For a self-propelled swimmer, there is no net force 
or torque, and therefore movement is a purely geometric process: the net displacement of a swimmer during a stroke is 
independent of the rate at which the stroke is executed, as long as the Reynolds and Strouhal numbers remain small enough.

Suppose that a swimmer occupies the closed compact domain $\Omega (t) \subset \mathbb{R}^3$, at time $t$, and
let $\partial \Omega (t)$ denote its prescribed time-dependent boundary. A \textit{swimming stroke} $\gamma$ is specified
by a time-dependent sequence of the boundary $\partial \Omega (t)$, and it is \textit{cyclic} if the initial and final shapes 
are identical, i.e.,~$\partial \Omega (0) = \partial \Omega (T)$ where $T$ is the period~\cite{shapere1989geometry}. 
The swimmer's boundary velocity $\mathbf{V}$ relative to fixed coordinates can be written as a part $\mathbf{v}$
that defines the intrinsic shape deformations, and a rigid motion $\mathbf{U}$. If $\mathbf{u}$ denotes the velocity
field in the fluid exterior to $\Omega$, then a standard LRN self-propulsion problem is:
\begin{quote}
Given a cyclic shape deformation $\gamma (t)$ that is specified by $\mathbf{v}$, solve the Stokes equations
subject to:
\begin{eqnarray*}
& & \textrm{Boundary condition:} \quad \mathbf{u} \big|_{\partial \Omega (t)} = \mathbf{V} = \mathbf{v} + \mathbf{U}, \quad  \mathbf{u} |_{\mathbf{x} \rightarrow \infty} = \mathbf{0} \\
& & \textrm{Force-free condition:} \quad \int_{\partial \Omega (t)} \mathbf{f} = \mathbf{0} \\
& & \textrm{Torque-free condition:} \quad \int_{\partial \Omega (t)} \mathbf{r} \wedge \mathbf{f} = \mathbf{0}
\end{eqnarray*}
\end{quote}

In order to treat general shape changes of a swimmer defined by $\Omega (t) \subset \mathbb{R}^3$ with boundary $\partial \Omega (t)$,
one must solve the exterior Stokes Equations (\ref{Eq.Stokes}) for $\mathbf{u}$, with a prescribed velocity $\mathbf{v} (t)$ on $\partial \Omega (t)$
and subject to the decay conditions $\mathbf{u} \sim 1/r$ and $p \sim 1 / r^2$ as $r \rightarrow \infty$.  The solution has the representation:
\begin{eqnarray*}
\mathbf{u} (\mathbf{x}) = - \dfrac{1}{8 \pi \mu} \int_{\partial \Omega (t)} \mathbf{G} (\mathbf{x}, \mathbf{y}) \cdot \mathbf{f} (\mathbf{y}) d S (\mathbf{y})
+ \dfrac{1}{8 \pi}  \int_{\partial \Omega (t)} \mathbf{v} (\mathbf{y}) \cdot \mathbf{T} (\mathbf{y}, \mathbf{x}) \cdot \mathbf{n} d S (\mathbf{y})
\end{eqnarray*}
where $\mathbf{G}$ is the free-space Green's function, $\mathbf{T}$ is the associated third-rank stress tensor,
$\mathbf{n}$ is the exterior normal and $\mathbf{f}$ is the force on the boundary~\cite{pozrikidis1992boundary}.
The constraints that the total force and the total torque vanish determine the center-of-mass translational and
angular velocities. When $\mathbf{x} \in \partial \Omega (t)$, this is an integral equation for the force distribution
on the boundary, the solution of which determines the forces needed to produce the prescribed shape changes. 

\section{Linked-Sphere LRN Swimmers\label{Sec.LinkSphBasic}}
To date, various simple linked-sphere models for which both analytical and computational results can be obtained have 
appeared. In particular, there are three linked-sphere models that are designed to investigate the effects of changing body length
vs. mass transportation on LRN swimming behavior. The three models are: the NG 3-sphere accordion model (Figure \ref{Fig.NG-PMPY-VE}a)
\cite{najafi2004simple, golestanian2008analytic, alexander2009hydrodynamics}, the PMPY 2-sphere model (Figure \ref{Fig.NG-PMPY-VE}b)
\cite{avron2005pushmepullyou}, and the VE 3-sphere model (Figure \ref{Fig.NG-PMPY-VE}c)~\cite{wang2012models}.

\subsection{Fundamental Solutions for Translating and Radially Deforming Spheres\label{Sec.FundSolnSph}} 

In either of the three models, shape deformations either come from linking-arm length change (the NG and the PMPY models) or
radial deformations of the spheres (the PMPY and the VE models).
When a sphere of radius $a(t)$ is pulled through a quiescent fluid with a steady force $\mathbf{f}$ under no-slip conditions at the
surfaces, the resulting flow field is given by~\cite{kim2013microhydrodynamics}:
\begin{eqnarray}\label{Eq.u-trans}
\mathbf{u} (\mathbf{r}) = \mathbf{f} \cdot \Big( 1 + \dfrac{a^2}{6} \nabla^2 \Big) \dfrac{\mathbf{G} (\mathbf{x}, \mathbf{x}_0)}{8 \pi \mu}
=\dfrac{\mathbf{f}}{8 \pi \mu r} \cdot \Big[ \mathbf{I} + \dfrac{\mathbf{r} \mathbf{r}}{r^2} + \dfrac{a^2}{3 r^2} \Big( \mathbf{I} - 3\dfrac{\mathbf{r} \mathbf{r}}{r^2} \Big) \Big]
\end{eqnarray}
where $\mathbf{x}_0$ is the position of the center of the sphere and $\mathbf{r} = \mathbf{x} - \mathbf{x}_0$, and the resulting 
velocity of the sphere is given by Stokes' law 
\begin{eqnarray}\label{Eq.StokesLaw}
\mathbf{f} = 6 \pi \mu a \mathbf{U}
\end{eqnarray}

A second fundamental solution is the velocity field $\mathbf{u}$ produced by a radially-expanding sphere, which can be generated by a
point source at the center $\mathbf{x}_0$ of the sphere. The corresponding velocity is~\cite{pozrikidis1992boundary}:
%
%
%
%
%
\begin{eqnarray}\label{Eq.u-RadDeform}
\mathbf{u} = \dot{a} \big( \dfrac{a}{r} \big)^2 \dfrac{\mathbf{r}}{r} = \dfrac{\dot{v}}{4 \pi r^2} \hat{\mathbf{r}}
\end{eqnarray}
where $v = 4 \pi a^3/3$ is the volume of the sphere and $\hat{\mathbf{r}} = \mathbf{r}/r$. 

By combining the two fundamental solutions (\ref{Eq.u-trans}) and (\ref{Eq.u-RadDeform}), we obtain the solution of the flow velocity
generated by a sphere of radius $a$, centered at $\mathbf{x}_0$, subject to a pulling/pushing force $\mathbf{f}$, and radially
deforming at the rate $\dot{a}$:
\begin{eqnarray*}
\mathbf{u} (\mathbf{r}; a, \mathbf{f}, \dot{a}) = \dfrac{1}{24 \pi \mu r}
\Big[ \Big( 3 + \dfrac{a^2}{r^2} \Big) \mathbf{f} + 3 \Big( 1- \dfrac{a^2}{r^2}  \Big) \big( \mathbf{f} \cdot \hat{\mathbf{r}} \big) \hat{\mathbf{r}} \Big]
+ \dot{a} \big( \dfrac{a}{r} \big)^2 \hat{\mathbf{r}}
\end{eqnarray*}

Next, the fundamental solution involves the interaction between two spheres. Suppose that at time $t$ the $i$th ($i=1,2$) sphere
has radius $a_i (t)$, centered at $\mathbf{x}_i (t)$, subjected to force $\mathbf{f}_i (t)$ and with radially deforming rate $\dot{a}_i (t)$.
In the scenario that the two spheres are far apart, i.e., $a_i / l \ll 1$, where~$ l = |\mathbf{x}_1 - \mathbf{x}_2|$,
the translational velocity of the $i$th sphere can be obtained by the \textit{Reflection Method}~\cite{kim2013microhydrodynamics,
yamakawa1970transport, wajnryb2013generalization, zuk2014rotne, liang2013fast, wang2018analysis}:
\begin{eqnarray}\label{Eq.U-Ref}
\mathbf{U}_i \sim \dfrac{\mathbf{f}_i}{6 \pi \mu a_i} + \delta \mathbf{U}_i^t \{ j \}+ \delta \mathbf{U}_i^e  \{ j \} + O  \Big( \dfrac{1}{l^4}  \Big) 
\end{eqnarray}
where $\delta \mathbf{U}_i^t $ is due to the translation of the other sphere, which arises from the flow given by Equation~(\ref{Eq.u-trans})
and is 
\begin{eqnarray}\label{Eq.Ut-Ref}
\delta \mathbf{U}_i^t  \{ j \}= \Big( 1 + \dfrac{a_i^2}{6} \nabla^2 \Big) \mathbf{u} (\mathbf{r}) \big|_{\mathbf{r} = \mathbf{x}_i - \mathbf{x}_j}
= \dfrac{1}{8 \pi \mu l} \Big[ \Big( 1 + \dfrac{a_i^2 + a_j^2}{3 l^2} \Big) \mathbf{I} + \Big( 1 - \dfrac{a_i^2 + a_j^2}{l^2} \Big) \hat{\mathbf{l}} \hat{\mathbf{l}} \Big] \mathbf{f}_j
\end{eqnarray}
where $\mathbf{l} = \mathbf{x}_i - \mathbf{x}_j$. The velocity $ \delta \mathbf{U}_i^e$ is resulted from a flow generated by the radial deformation
of the other sphere, defined by Equation (\ref{Eq.u-RadDeform}):
\begin{eqnarray}\label{Eq.Ue-Ref}
\delta \mathbf{U}_i^e  \{ j \} = \Big( 1 + \dfrac{a_i^2}{6} \nabla^2 \Big) \mathbf{u} (\mathbf{r}) \big|_{\mathbf{r} = \mathbf{x}_i - \mathbf{x}_j}
 = \dot{a}_j \Big( \dfrac{a_j}{l} \Big)^2 \hat{\mathbf{l}}
\end{eqnarray}

Finally, the power consumption $P $ of a sphere with radius $a $, dragged by a force $\mathbf{f} $, translating at velocity $\mathbf{U}$ and 
radially expanding/contacting at a rate of $\dot{a} $ comprises two parts: $P^t = \mathbf{f} \cdot \mathbf{U}$ that results from
the drag force on the sphere, and $P^e = 16 \pi \mu a \dot{a}^2$ that results from the radial deforming~\cite{avron2005pushmepullyou, wang2018analysis}. 
Therefore the total power expended is:
\begin{eqnarray}\label{Eq.Power}
P = P^t + P^e = \mathbf{f} \cdot \mathbf{U} + 16 \pi \mu a \dot{a}^2
\end{eqnarray}

\subsection{Non-Dimensionalization of the System}

Let $A$, $L$ be the characteristic sphere radius and link-arm length. The reflection method holds in the regime
$\varepsilon := A/L \ll1$. Shape deformations of the linked-sphere models (see Sections \ref{Sec.NG}--\ref{Sec.VE}) come from 
the expanding/contracting of the link-arms ($\dot{l}$), or the radial expanding/contracting of the spheres ($\dot{a}$).
For the shape deformation scales, we require that $\dot{a}/\dot{l} \sim O (1)$, and $|\delta a |, |\delta l | \ll A$,
where $\delta a$ and $\delta l$ denote the shape deformations in $a$ and $l$, respectively.

We non-dimensionalize the radii and rod lengths by $A$ and $L$:
\begin{eqnarray*}
a^* (t) = \dfrac{a (t)}{A}, \quad l^* (t) = \dfrac{l (t)}{L}
\end{eqnarray*}

We non-dimensionalize other length scales by $A$ as well, and time by $T$-the swimming stroke period, thus
we obtain:
\begin{eqnarray*}
& &\mathbf{x}^* = \dfrac{\mathbf{x}}{A}, \ \nabla^* = A \nabla, \ t^* = \dfrac{t}{T}, \  \mathbf{u}^* = \dfrac{T}{A} \mathbf{u}, \ \mathbf{U}^* = \dfrac{T}{A} \mathbf{U} \\
& &\dot{a}^* = \dfrac{d a^*}{d t^*} = \dfrac{T}{A} \dot{a} , \qquad
\dot{l}^* = \dfrac{d l^*}{d t^*} =\dfrac{T}{A} \dot{l} 
\end{eqnarray*}

We assume that the deformations $\delta a, \delta l$ in $a$ and $l$ are of the same order, 
both of which are scaled by $A$,~thus 
\begin{eqnarray*}
\dot{a}^* = \lim_{ \delta t \rightarrow 0} \dfrac{\delta a}{\delta t} = 
 \lim_{ \delta t \rightarrow 0} \dfrac{T}{A} \dfrac{\delta a^*}{\delta t^*} = \dfrac{T}{A} \dot{a}, \qquad
 \dot{l}^* = \lim_{ \delta t \rightarrow 0} \dfrac{\delta l}{\delta t} = 
 \lim_{ \delta t \rightarrow 0} \dfrac{T}{A} \dfrac{\delta l^*}{\delta t^*} = \dfrac{T}{A} \dot{l}
\end{eqnarray*}
and we have the approximation:
\begin{eqnarray*} 
l^* (t^* + \delta t^*) = \dfrac{l (t + \delta t)}{L} \sim \dfrac{l (t) + \dot{l} (t) \delta t}{L} = l^* (t^*) + \varepsilon \dot{l}^* (t^*) \delta t^*
\end{eqnarray*}

The drag force $\mathbf{f}$ exerted on a sphere of radius $a$ in a quiescent fluid is related to 
the sphere velocity $\mathbf{U}$ via Equation (\ref{Eq.StokesLaw}), which leads to the following scaling
for forces:
\begin{eqnarray*}
\mathbf{f}^* = \dfrac{T}{6 \pi \mu A^2} \mathbf{f} = a^* \mathbf{U}^*
\end{eqnarray*}

Therefore Equations (\ref{Eq.U-Ref})--(\ref{Eq.Ue-Ref}) become:
\begin{eqnarray*}
\mathbf{U}_i^* &\sim& \dfrac{\mathbf{f}_i^*}{a_i^*} +  \delta \mathbf{U}_i^{t*}   \{ j \} + \delta \mathbf{U}_i^{e*}  \{ j \} + O (\varepsilon^4) \\
\delta \mathbf{U}_i^{t*}  \{ j \} &=& \dfrac{3 \varepsilon}{4 l^*} \Big[ \Big( 1 + \varepsilon^2 \dfrac{a_i^{*2} + a_j^{*2}}{3 l^{*2}} \Big) \mathbf{I} + 
\Big( 1 - \varepsilon^2 \dfrac{a_i^{*2} + a_j^{*2}}{l^{*2}} \Big) \hat{\mathbf{l}}^* \hat{\mathbf{l}}^* \Big] \mathbf{f}_j^* \\
\delta \mathbf{U}_i^{e*}  \{ j \} &=& \varepsilon^2 \dot{a}_j^* \Big( \dfrac{a_j^*}{l^*} \Big)^2 \hat{\mathbf{l}}^*
\end{eqnarray*}

Finally, we non-dimensionalize the power $P$ as
\begin{eqnarray*}
P^* = \dfrac{1}{6 \pi \mu} \dfrac{T^2}{A^3} P
\end{eqnarray*}
so that while $\textrm{Power} = \textrm{Force} \cdot \textrm{Velocity}$ in dimensional form, after non-dimensionalization we still have
$P^* = \mathbf{f}^* \cdot \mathbf{U}^*$. Thus the non-dimensional version of Equation (\ref{Eq.Power}) is:
\begin{eqnarray*}
P^* = \mathbf{f}^* \cdot \mathbf{U}^* + \dfrac{8}{3} a^* \dot{a}^{*2}
\end{eqnarray*}
%

\subsection{NG 3-Sphere Swimmer\label{Sec.NG}}
The NG swimmer (Figure \ref{Fig.NG-PMPY-VE}a) consists of three spheres with fixed radii $a_i \ (i=1,2,3)$ and two linking-arms
with adjustable length $l_i (t) \ (i=1,2)$~\cite{najafi2004simple, golestanian2008analytic, alexander2009hydrodynamics}. While the 
spheres are rigid, the linking-arms can stretch or contract. The geometry of the model is shown in Figure \ref{Fig.NG-PMPY-VE}a,
where the three spheres align along the $x$-axis, numbered from sphere 1 to 3 from left to right. Let $\mathbf{e} $ be the unit
vector pointing in the positive $x$ direction, and we have $\mathbf{e} = ( \mathbf{x}_i - \mathbf{x}_j)/\| \mathbf{x}_i - \mathbf{x}_j \|$
for any pair of $i,j$ with $i>j$. Thus we have $\mathbf{f}_i = f_i \mathbf{e}$ and $\mathbf{U}_i = U_i \mathbf{e}$ for $i=1,2,3$. For simplicity, we consider the simple 
geometry $a_1 = a_2 = a_3 =A$. Assume initial shape $l_1 (0) = l_2 (0) = L$.

The (non-dimensional) velocity of each sphere is given by:
\begin{eqnarray*}
U_i^* = f_i^*  +  \sum_{j \neq i} \delta U_i^{t*}   \{ j \} = f_i^*  +  \sum_{j \neq i}  \dfrac{3 \varepsilon}{2 l^* (i,j)} \Big( 1 - \dfrac{2}{3} \varepsilon^2  \dfrac{1}{l^{*2} (i,j)} \Big) f_j
\end{eqnarray*}
where $l^* (i,j)$ is the scaled distance between spheres $i$ and $j$. The velocities are related via the following~relations:
\begin{eqnarray*}
U_2^* - U_1^* = \dot{l}_1^*, \qquad U_3^* - U_2^* = \dot{l}_2^*
\end{eqnarray*}

The system is force-free,
\begin{eqnarray*}
\sum_{i=1}^3 f_i^*  = 0
\end{eqnarray*}
ad the velocity and power of the swimmer are: 
\begin{eqnarray*}
U^* = \dfrac{1}{3} \sum_{i=1}^3 U_i^*, \quad P^* = \sum_{i=1}^3 f_i^* U_i^*
\end{eqnarray*}
%

\subsection{PMPY 2-Sphere Swimmer\label{Sec.PMPY}}
The PMPY swimmer (Figure \ref{Fig.NG-PMPY-VE}b) consists of two spheres with radii $a_i (t) \ (i=1,2)$ and one linking-arm
with length $l (t)$~\cite{avron2005pushmepullyou}. The spheres can expand or contract in the radial direction, and the linking-arm can stretch or contract. 

The (non-dimensional) velocity of each sphere is given by:
\begin{eqnarray*}
U_i^* &=& \dfrac{f_i^*}{a_i^*}  +  \big[ \delta U_i^{t*}   \{ j \} + \delta U_i^{e*}   \{ j \} \big]_{j \neq i} \\
&=& \dfrac{f_i^*}{a_i^*}  +  \Big[ \dfrac{3 \varepsilon}{2 l^*} \Big( 1 - \dfrac{1}{3} \varepsilon^2  \dfrac{a_i^{*2} + a_j^{*2}}{l^{*2}} \Big) f_j 
+ (-1)^{\textrm{sign} (i-j)} \varepsilon^2 \dot{a}_j^* \Big( \dfrac{a_j^*}{l^*} \Big)^2    \Big]_{j \neq i}
\end{eqnarray*}

The velocities are related by:
\begin{eqnarray*}
U_2^* - U_1^* = \dot{l}^*
\end{eqnarray*}
and the force-free condition is: 
\begin{eqnarray*}
\sum_{i=1}^2 f_i^* = 0
\end{eqnarray*}

Assume initial conditions $a_1 (0) = a_2 (0) = A$ and $l (0) = L$.
There is no mass exchange between the swimmer and surrounding fluid, thus the total volume of the two spheres is conserved:
\begin{eqnarray*}
a_1^{*3} + a_2^{*3}  = 2, \qquad \sum_{i=1}^2 a_i^{*2} \dot{a}_i^* = 0
\end{eqnarray*}

The velocity and power of the whole swimmer are: 
\begin{eqnarray*}
U^* = \dfrac{1}{2} \sum_{i=1}^2 U_i^*, \quad P^* =   \sum_{i=1}^2 \big(  f_i^* U_i^* + \dfrac{8}{3} a_i^* \dot{a}_i^{*2} \big)
\end{eqnarray*}
%

\subsection{VE 3-Sphere Swimmer\label{Sec.VE}}
The VE swimmer (Figure \ref{Fig.NG-PMPY-VE}c) consists of three spheres with radii $a_i (t) \ (i=1,2,3)$ and two rigid linking-arms with
fixed length $l_i \ (i=1,2)$~\cite{wang2012models}. The spheres can only expand or contract in the radial direction, and each can only
exchange mass with its neighboring sphere(s), that is, between sphere 1 and 2, or between sphere 2 and 3, but not between sphere
1 and 3.

The (non-dimensional) velocity of each sphere is given by:
\begin{eqnarray*}
U_i^* &=& \dfrac{f_i^*}{a_i^*}  + \sum_{j \neq i}  \big[ \delta U_i^{t*}   \{ j \} + \delta U_i^{e*}   \{ j \} \big]  \\
&=& \dfrac{f_i^*}{a_i^*}  +  \sum_{j \neq i}  \Big[ \dfrac{3 \varepsilon}{2 l^* (i,j)} \Big( 1 - \dfrac{1}{3} \varepsilon^2  \dfrac{a_i^{*2} + a_j^{*2}}{l^{*2}(i,j)} \Big) f_j 
+ (-1)^{\textrm{sign} (i-j)} \varepsilon^2 \dot{a}_j^* \Big( \dfrac{a_j^*}{l^*(i,j)} \Big)^2    \Big] 
\end{eqnarray*}

Assume initial conditions $a_1 (0) = a_2 (0) = a_3 (0) = A$ and $l_1 = l_2 = L$.
The system is force-free and total volume is conserved:
\begin{eqnarray*}
\sum_{i=1}^3 f_i^* = 0, \qquad  \sum_{i=1}^3 a_i^{*3} = 3  , \qquad  \sum_{i=1}^3 a_i^{*2} \dot{a}_i^* = 0
\end{eqnarray*}

The velocity and power of the whole swimmer are: 
\begin{eqnarray*}
U^* = U^*_1 = U^*_2 = U^*_3, \quad P^* =   \sum_{i=1}^3 \big(  f_i^* U_i^* + \dfrac{8}{3} a_i^* \dot{a}_i^{*2} \big)
\end{eqnarray*}

Notice that the linking-arms are rigid.

\section{Optimal Strokes of NG, PMPY and VE Swimmers\label{Sec.OptBasic}}\unskip

\subsection{Euler--Lagrange Equation for Optimal Strokes of LRN Swimmers}\label{Sec.EL-SM}

({{Hereafter we will continue} our discussion using the non-dimensional quantities but omitting 
the $*$ notation for simplicity).} 
As is introduced in Section \ref{Sec.IntroSwim}, a swimming stroke $\gamma (t)$ is a $t$-series of swimmer shapes. In the case
when the swimmer has only finitely many degrees of freedom in their shape deformations, 
$\gamma$ can be considered as a function from $[0, T]$ to $\mathbb{R}^m$,
where $m$ is the number of degrees of freedom of the swimmer, and each component in $\gamma (t)$ defines one degree of freedom.
For~example, all three linked-sphere swimmers in Sections \ref{Sec.NG}--\ref{Sec.VE} have only 2 degrees of freedom, thus~we have $\gamma_{\textrm{NG}} (t) = (l_1 (t), l_2 (t))^T$ for a NG swimmer, $\gamma_{\textrm{PMPY}} (t) = (l (t), a_1 (t))^T$ for a 
PMPY swimmer, and $\gamma_{\textrm{VE}} (t) = (a_1 (t), a_3 (t))^T$ for a VE swimmer.

The linear and driftless properties of Stokes flows make the LRN swimming problem a classic driftless controllable system 
\cite{shapere1989geometry, cherman2000low, loheac2013controllability, loheac2014controllability, chambrion2019optimal, avron2004optimal, leshansky2007frictionless}.
Consider a cyclic stroke $\gamma (t)$ of a LRN swimmer, where $  t \in [0,1], \ \gamma (0) = \gamma (1)$. The net translation $X (\gamma)$ and
energy dissipation $\mathcal{E} (\gamma)$ within a stroke can be represented as:
\begin{eqnarray}\label{Eq.NetTr-X}
X (\gamma) &=& \int_0^1 U (\gamma) dt = \int_0^1 \mathcal{F} (\gamma) \cdot \dot{\gamma} dt \\ \label{Eq.Energy-E}
\mathcal{E} (\gamma) &=& \int_0^1 P (\gamma) dt = \int_0^1 \big( \mathcal{G} (\gamma) \dot{\gamma},  \dot{\gamma} \big) dt
\end{eqnarray}
where for NG, PMPY and VE equations, the operators $\mathcal{F}$ and $\mathcal{G}$ can be calculated from the
equations from Sections \ref{Sec.NG}--\ref{Sec.VE}, respectively.
For a given initial shape which we denote by $\gamma_0$ and a given net translation $X_0$,  
let $\Gamma \{ \gamma_0,X_0 \}$ be the set of all cyclic strokes that starts and 
ends with shape $\gamma_0$ and results in a net translation of $X_0$ in a cycle:
\begin{eqnarray*}
\Gamma \{ \gamma_0 , X_0 \} = \{ \gamma \big| \gamma (0) = \gamma (1) = \gamma_0, X (\gamma) =X_0 \}
\end{eqnarray*}

The optimal stroke problem can be formulated as follows:
\begin{quote}
Given an initial shape $\gamma_0$ and a net translation $X_0$, find the stroke $\overline{\gamma} $ in $\Gamma \{ \gamma_0, X_0 \} $ that
minimizes the energy dissipation $\mathcal{E}$:
\begin{eqnarray}\label{Eq.OptStr}
\mathcal{E} (\overline{\gamma}) = \min_{ \gamma \in \Gamma \{ \gamma_0 , X_0 \} } \mathcal{E} (\gamma)
\end{eqnarray}
\end{quote}

The Euler--Lagrange equation for this optimization problem is~\cite{alouges2008optimal, alouges2009optimal}:
\begin{eqnarray}\label{Eq.EL}
-\dfrac{d}{d t} \big( \mathcal{G} \dot{\gamma} \big) + \dfrac{1}{2}
\left(
\begin{array}{c}
(\partial_{\gamma_1} \mathcal{G} \dot{\gamma}, \dot{\gamma}) \\
(\partial_{\gamma_2} \mathcal{G} \dot{\gamma}, \dot{\gamma}) \\
\vdots \\
(\partial_{\gamma_m} \mathcal{G} \dot{\gamma}, \dot{\gamma}) \\
\end{array}
\right)
+ \lambda \big[  (\nabla \mathcal{F})^T - \nabla \mathcal{F}\big] \cdot \dot{\gamma} = 0
\end{eqnarray}
where $m$ is the number of degrees of freedom in the swimmer's shape deformation, and $\gamma = (\gamma_1, \gamma_2, \cdots, \gamma_m)^T \in \mathbb{R}^m$.
{ Solutions of Equation (\ref{Eq.EL}) give the geodesics connecting $(\gamma_0, 0)$ and $(\gamma_0, X_0)$,
which we refer to as the geodesic strokes. The optimal stroke is the geodesic stroke that consumes the least energy~\cite{alouges2008optimal, alouges2009optimal}.}

We use the \textit{Shooting Method} (SM) to numerically solve the governing Equation (\ref{Eq.EL}) for the
optimal stroke problem. The numerical methods are
developed in~\cite{alouges2008optimal, alouges2009optimal} and have been applied for NG { with equal sizes} and PMPY swimmers;
{ recent research works also report the efficiency of NG swimmers with unequal sized sphere~\cite{nasouri2019efficiency}.}
In the following we use SM for all three (NG, PMPY and VE) linking-arm swimmers and compare their swimming performances.
In Appendix \ref{Sec.SM} we present an outline of SM for the convenience of the readers of this paper.

\subsection{Numerical Results\label{Sec.CompNum_Org}}

\begin{figure}[h]
\centering
\includegraphics[width=\textwidth]{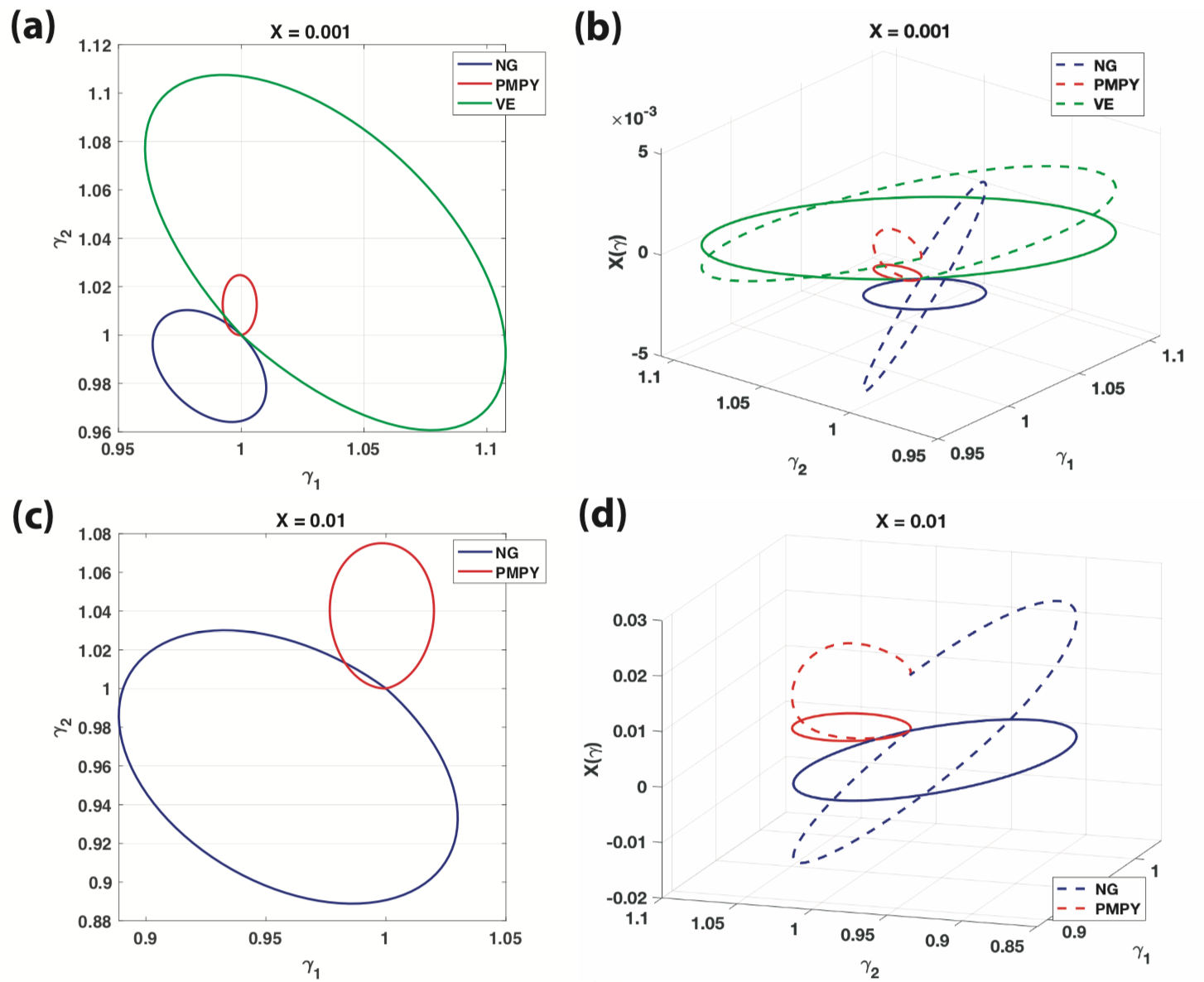}
\caption{\textbf{{Optimal strokes of NG 3-sphere, PMPY 2-sphere and VE 3-sphere swimmers.}} 
({\textbf{a,c}}) give the $\gamma$-paths of the optimal strokes
of the NG (blue), PMPY (red) and VE (green) swimmers, when~$X(\gamma) = 0.001, \ 0.01$, respectively.
({\textbf{b,d}}) give the optimal stroke trajectories $X( \overline{\gamma} (t))$, $t \in [0,1]$ (dashed lines) corresponding to the $\gamma$-paths (solid lines).
$(\gamma_1, \gamma_2 ) = (l_1, l_2)$ for NG swimmer; $(\gamma_1, \gamma_2 ) = (l, a_1)$ for PMPY;
$(\gamma_1, \gamma_2 ) = (a_1, a_3)$ for VE. All strokes start from $(\gamma_1 (0), \gamma_2 (0) ) = (1,1)$.
}
\label{Fig.Compare-NPV}
\end{figure}

The numerical results of optimal strokes of a NG 3-sphere, PMPY 2-sphere and VE 3-sphere swimmers are given in Table \ref{Tab.E-NPV}
and Figure \ref{Fig.Compare-NPV}, where we take $\varepsilon = A/L = 0.2$ in all simulations. For~different values of the net translation
$X $, the optimal strokes $\overline{\gamma}$ of the three swimmers are computed by the SM method described in Section \ref{Sec.EL-SM} and Appendix \ref{Sec.SM}, the 
energy dissipation of the optimal strokes $\mathcal{E} (\overline{\gamma})$ is recorded in Table \ref{Tab.E-NPV}, and the $\gamma$-paths
(solid lines) and the $X( \overline{\gamma} (t))$, $t \in [0,1]$ trajectories (dashed lines) are given in Figure \ref{Fig.Compare-NPV}.
The simulation results show that PMPY is the most efficient swimmer among the three, next is NG, the last is VE:
\begin{eqnarray*}
\textrm{Eff}_{\textrm{PMPY}} > \textrm{Eff}_{\textrm{NG}} \sim \textrm{Eff}_{\textrm{VE}}
\end{eqnarray*}
where we define the \textit{efficiency} of an LRN swimmer as the net translation to energy dissipation ratio of the optimal stroke of the swimmer:
\begin{eqnarray}\label{Eq.DefEff}
\textrm{Eff} = \dfrac{X (\overline{\gamma})}{\mathcal{E} (\overline{\gamma})}, \qquad \textrm{where $\overline{\gamma}$ is the optimal stroke.}
\end{eqnarray}

The dimensional form of Eff has the unit of force$^{-1}$. In addition, from Table \ref{Tab.E-NPV} we find that { for single loop optimal strokes}, 
the efficiency for each swimmer is about the same for different values of $X (\overline{\gamma})$, { with a slight increase in the efficiency for NG and PMPY swimmers
as $X (\overline{\gamma})$ increases}: 
\begin{eqnarray*}
\textrm{Eff}_{\textrm{PMPY}} \sim  29.5 - 33.0 \times 10^{-3} ,  \quad 
\textrm{Eff}_{\textrm{NG}} \sim 2.6 - 3.1 \times 10^{-3},  \quad 
\textrm{Eff}_{\textrm{VE}} \sim 1 \times 10^{-3}
\end{eqnarray*}

As $X$ increases, { geodesic} strokes of multiple loops show up for NG and PMPY swimmers. However, we do not catch multi-loop solutions in our 
numerical simulations for VE swimmers, as~large shape deformations of VE models easily lead to negative volume of the spheres. Comparing to single loop 
{ geodesic} strokes, multi-loop { geodesic} strokes are less efficient (see { Appendix \ref{Sec.GeoStroke}}).  

We summarize our main conclusions from the above numerical results as follows:
\begin{enumerate}[leftmargin=*,labelsep=4.9mm]
\item A PMPY model that adopts a mixed-mode of shape deformations is the most efficient among the three; next is NG and the last is VE, both
of which adopt a single-mode of shape deformations (see Remark below). 
\item Single-loop { geodesic} strokes are more efficient than multi-loop { geodesic} strokes.
\item The efficiency of a given LRN swimmer is almost the same for different $X$, as long as the optimal stroke is single-looped.
\end{enumerate}

{Finally, we numerically compare the optimal strokes of the three model swimmers to stepwise square strokes. The results
are presented in Appendix \ref{Sec.Comp-OptSq}, which clearly shows the improvement of swimming efficiency in the optimal strokes
comparing to the stepwise square strokes.
}

\begin{table}
\caption{{Energy dissipation} $\mathcal{E} (\overline{\gamma})$ and efficiency Eff.} 
\centering
\begin{tabular}{ccccccc}
\hline
\boldmath{$X (\overline{\gamma})$}	& \multicolumn{2}{c}{\textbf{NG 3-Sphere}} 	&  \multicolumn{2}{c}{\textbf{PMPY 2-Sphere}} & \multicolumn{2}{c}{\textbf{VE 3-Sphere} } \\\hline
   & \boldmath{$\mathcal{E} (\overline{\gamma})$} & \textbf{Eff}   & \boldmath{$\mathcal{E} (\overline{\gamma})$} & \textbf{Eff}    & \boldmath{$\mathcal{E} (\overline{\gamma})$} & \textbf{Eff}    \\
\hline
$0.001$	& $37.61 \times 10^{-2} $  &  $ 2.66 \times 10^{-3} $ &$ 3.37 \times 10^{-2}$ & $29.67 \times 10^{-3}$	&$ 95.87 $  & $1.04 \times 10^{-3}$ \\
 (Figure \ref{Fig.Compare-NPV}a,b)             &  &      &  \\   
 &  &      &  \\   
$0.005$        & $182.96 \times 10^{-2}$  &  $2.73 \times 10^{-3}$  & $16.82 \times 10^{-2}$  &  $29.73 \times 10^{-3}$ & $539.33 \times 10^{-2}$ & $0.93 \times 10^{-3}$ \\      
              &  &      &  \\   
$0.01$       & $358.28 \times 10^{-2}$ & $2.79 \times 10^{-3}$   & $33.55 \times 10^{-2}$  &  $29.81 \times 10^{-3}$ & - & - \\   
 (Figure \ref{Fig.Compare-NPV}c,d)      &  &      &  \\   
              &  &      &  \\   
$0.02$       & $695.46  \times 10^{-2}$  & $2.88 \times 10^{-3}$   & $66.77 \times 10^{-2}$ & $29.95 \times 10^{-3}$  & - & - \\      
                            &  &      &  \\   
$0.05$       &  $1636.25 \times 10^{-2}$   & $3.06 \times 10^{-3}$ &  $ 164.91 \times 10^{-2}$ & $30.33 \times 10^{-3}$ & - & - \\    
                 &  &      &  \\   
$0.1$       & -  & -   & $320.94  \times 10^{-2}$ & $31.16 \times 10^{-3}$ & -& -  \\     
                            &  &      &  \\   
$0.2$       & -  &  -  & $606.30  \times 10^{-2}$ & $32.99 \times 10^{-3}$ & - & -  \\     
\hline
\end{tabular}
\label{Tab.E-NPV}
\end{table}

\newtheorem*{remark}{Remark}

\begin{remark}
{At LRN, different microorganisms adopt various propulsion mechanisms and directed locomotion strategies for searching for food and
running from predators. While many microorganisms including bacteria use a flagellated or ciliated mode of swimming~\cite{taylor1952action, 
hancock1953self, higdon1979hydrodynamics, phan1987boundary, lauga2009hydrodynamics, elgeti2015physics, sokolov2010swimming, 
riedel2005self, lauga2006swimming, rafai2010effective, patteson2015running, sowa2008bacterial}, some cells use an 
amoeboid swimming strategy, which~relies on the generation, protrusion and even travel of pseudopodia or blebs 
\cite{barry2010dictyostelium, bae2010swimming, van2011amoeboid}. Such shape deformations propagating over the 
whole cell body can usually be generalized as a combination of two modes: stretching of the cell body, and mass transportation
along the cell body. Comparing with the linked-sphere types of models, cell body stretching can be considered as mode $\dot{l}$ while 
mass transportation as mode $\dot{a}$. According to the \textit{scallop theorem}~\cite{purcell1977life}, a minimal LRN swimmer has
at least two degrees of freedom in its shape deformations, therefor we have three possible combinations: two single-mode of shape 
changes: both in body stretching (both are $\dot{l}$-type of shape deformations), both in mass transportation (both are $\dot{a}$-type of shape
deformations), and a mixed-modes of shape change: body stretching combined with mass transportation $(\dot{l}, \dot{a})$. It is clearly seen that the three combinations
correspond to the three linked-sphere swimmers (NG, VE and PMPY) we discussed here, and our numerical results indicate that:
\begin{quote}
Comparing to single-mode shape deformations, mixed-modes of shape deformations, i.e.,~body stretching combined with mass transportation
is more efficient when swimming at~LRN.
\end{quote}
An asymptotic analysis study on the three swimmers that derives the same conclusion can be found in~\cite{wang2015performance},
which illustrates that a PMPY swimmer has a net translation $X (\gamma) \sim O (1)$, comparing to $X (\gamma) \sim O (\varepsilon^2)$
for a NG or a VE swimmer. Therefore the mixed control strategy as is adopted by a PMPY swimmer is superior to uni-mode of control
strategies as are adopted by a NG or a VE swimmer.
}
\end{remark}

\begin{remark}
{A more popular definition of efficiency of NG swimmer in its dimensional form is given  by~\cite{lighthill1952squirming}:
\begin{eqnarray*}
\textrm{Eff}^{-1} = \dfrac{\frac{1}{T} \int_0^T P (\gamma) dt}{6 \pi \mu A X^2 }
\end{eqnarray*}
which measures the ratio of the average energy dissipation in the stroke $\gamma$ to that applied to a rigid sphere of radius $A$ traveling
the same distance $X$. However, this definition does not apply well to our discussion in that (1) we are considering non-rigid spheres where
the spheres are allowed to deform radially, and (2) we are comparing swimmers of different numbers of spheres. Therefore we adopt the definition
given by Equation (\ref{Eq.DefEff}) instead.
}
\end{remark}


\section{Optimal Strokes of \boldmath{$(m+1)$}-Linked-Sphere LRN Swimmers\label{Sec.OptChain}}

In this section we will discuss $(m+1)$-linked-sphere LRN swimmers, starting from the NG-type of swimmers (Section \ref{Sec.NG-NSph}),
where rigid spheres are considered and shape deformations are only allowed to be in linking-arms length changes. 
Next we consider the PMPY-type of swimmers (Section \ref{Sec.PMPY-NSph}), where both spheres and linking-arms are allowed to deform. 
Finally we discuss the effect of sphere separation distance on the swimmer's swimming efficiency (Section \ref{Sec.NG-PMPY-FarApartSph}).

\subsection{$(m+1)$-Linked-Sphere NG Swimmers\label{Sec.NG-NSph}}

\begin{figure}[h]
\centering
\includegraphics[width=0.7\textwidth]{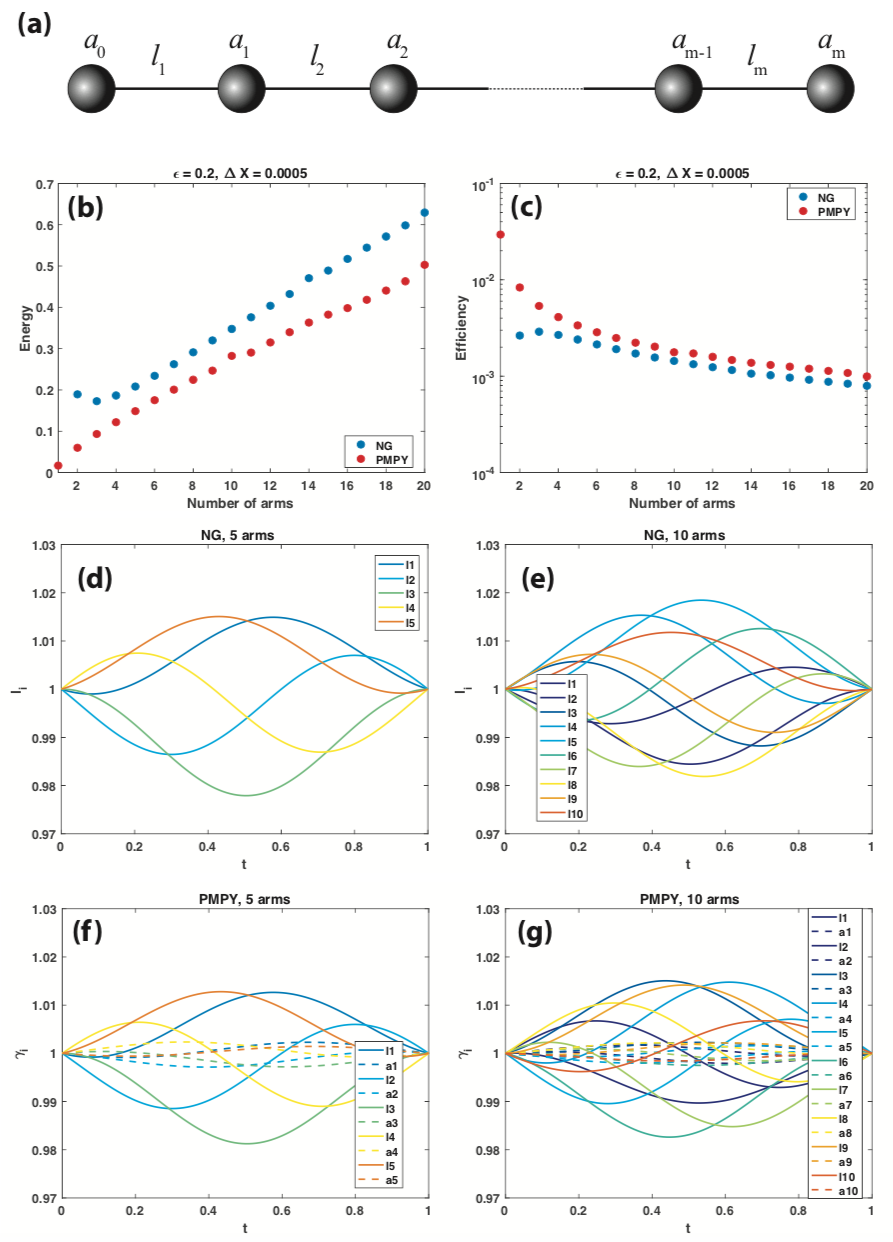}
\caption{\textbf{$(m+1)${-linked-sphere NG-type or PMPY-type of swimmers.}} (\textbf{a}) Cartoon illustration of the swimmer, which consists of
$m+1$ spheres and $m$ linking-arms. (\textbf{b}) Energy dissipation of optimal strokes 
$\mathcal{E} (\overline{\gamma})$ and (\textbf{c}) efficiency Eff of NG (blue dots) and PMPY (red dots) swimmers with $m$ linking-arms. 
(\textbf{d},\textbf{e}) The $\gamma$-path of the optimal stroke $\overline{\gamma}$ of a NG swimmer with 5 and 10 linked arms, respectively,
where $\gamma = (l_1, l_2, \cdots, l_m)^T$. (\textbf{f},\textbf{g}) The $\gamma$-path of the optimal stroke $\overline{\gamma}$ of a PMPY swimmer with 5 and 10 linked arms, respectively,
where $\gamma = (l_1, l_2, \cdots, l_m, a_1, a_2, \cdots, a_m)^T$, where the solid lines give the $l_i$-trajectories
and dashed lines the $a_i$-trajectories. $\varepsilon = 0.2$ and $X = 5 \times 10^{-4}$ in all simulations.
}
\label{Fig.N-link-NG}
\end{figure}

An $(m+1)$-linked-sphere NG swimmer consists of $m+1$ spheres with fixed radii $a_i \ (i=0,1,2,\cdots, m)$ and $m$ linking-arms 
with adjustable length $l_i (t) \ (i=1,2, \cdots, m)$. The geometry of the model is shown in Figure \ref{Fig.N-link-NG}a, where the
swimmer lies along the $x$-axis, and the spheres and linking-arms are numbered from left to right. We assume equal-sized 
spheres $a_0 = a_1 = \cdots = a_m = A$ and initial shape $l_1 (0) = l_2 (0) = \cdots = l_m (0) = L$. The equation system
describing the mechanics of the model follows that in Section \ref{Sec.NG}. We follow the same numerical scheme to solve 
the optimization problem Equation (\ref{Eq.OptStr}) for different numbers of linking-arms $m$.
In our simulations, we test for $m \in [2,20]$ and with small net translation $X = 5 \times 10^{-4}$ and $\varepsilon = A/L = 0.2 $ in all simulations.

The energy dissipation of the optimal stroke reaches a minimum $\min_{m} \mathcal{E} (\overline{\gamma}) = 17.28 \times 10^{-2}$ at $m = 3$, i.e., 
a 4-sphere NG swimmer; after that, $\mathcal{E} (\overline{\gamma}) $ increases as $m$ increases (Figure \ref{Fig.N-link-NG}b, blue dots).
Correspondingly, the most efficient swimmer is the 4-sphere NG swimmer with Eff $=2.89 \times 10^{-3}$, and~the efficiency decreases as $m$ increases 
(Figure \ref{Fig.N-link-NG}c, blue dots).

According to the \textit{scallop theorem}~\cite{purcell1977life}, a minimal LRN swimmer should have at least 2 degrees of freedom.
When it comes to LRN swimmers with more than 2 degrees of freedom, the question is: is it better to use all degrees of freedom or
to take a subset of them so to reach more efficient swimming strategy? In our simulations, we find that the optimal strokes for different
numbers of linking-arms have all made use of all degrees of freedom, for example, Figure \ref{Fig.N-link-NG}d,e show the $\gamma$-paths
of the optimal strokes for 5 and 10 arms, respectively, which show that all linking-arms performed length deformations in the optimal stroke, none
is disabled.

\subsection{$(m+1)$-Linked-Sphere PMPY Swimmers\label{Sec.PMPY-NSph}
}
Next we consider $(m+1)$-linked-sphere PMPY swimmers, which consist of $m+1$ spheres with adjustable 
radius $a_i (t) \ (i=0, 1,2, \cdots, m)$ and $m$ linking-arms with adjustable length $l_i (t) \ (i=1,2, \cdots, m)$
(Figure \ref{Fig.N-link-NG}a). We assume initial shape $a_0 (0) = a_1 (0) = \cdots = a_m (0) = A$ and $l_1 (0) = l_2 (0) = \cdots = l_m (0) = L$.
Under the volume conservation constraint of all spheres, we take $a_1, a_2, \cdots, a_m$ to be controls while $a_0(t)$ can be 
obtained from the volume conservation constraint. The equation system of the model follows that in Section \ref{Sec.PMPY}. 
In our simulations, we test for $m \in [1,20]$ and with small net translation $X = 5 \times 10^{-4}$ and $\varepsilon = A/L = 0.2 $ in all simulations.

The energy dissipation of the optimal stroke reaches a minimum $\min_{m} \mathcal{E} (\overline{\gamma}) = 1.70 \times 10^{-2}$ at $m = 1$, i.e., 
a 2-sphere PMPY swimmer, which is the one we discussed in Section \ref{Sec.PMPY}. $\mathcal{E} (\overline{\gamma}) $ increases as $m$ increases (Figure \ref{Fig.N-link-NG}b, red dots).
Correspondingly, the most efficient swimmer is the 2-sphere NG swimmer with Eff $=29.5 \times 10^{-3}$, and the efficiency decreases as $m$ increases 
(Figure \ref{Fig.N-link-NG}c, red dots). Comparing to NG type of swimmers with the same structure (i.e., same numbers of spheres and
linking-arms), PMPY swimmers consumes less energy and is thus more efficient (Figure \ref{Fig.N-link-NG}b,c).

Similar to NG swimmers, for PMPY swimmers, the optimal strokes for different numbers of linking-arms have also made use of all degrees of freedom.
Figure \ref{Fig.N-link-NG}fg show the $\gamma$-paths of the optimal strokes for 5 and 10 arms, respectively, which show that all linking-arms performed length 
deformations and all spheres performed radial deformations in the optimal stroke. In addition, we find that for large $m$, the amplitude
of sphere radial deformations $a_i (t)$ are smaller than that of the linking-arms length changes $l_i (t)$ (Figure \ref{Fig.N-link-NG}f,g); on the 
other hand, in the most efficient PMPY type of swimmers, i.e., the 2-sphere PMPY swimmer, the amplitude of sphere radial deformation $a_1(t)$ is larger than
that of the linking-arm length change $l (t)$ (Figure \ref{Fig.Compare-NPV}a,c).

\subsection{$(m+1)$-Linked-Sphere NG and PMPY Swimmers With Widely Separated Spheres\label{Sec.NG-PMPY-FarApartSph}}

\begin{figure}[h]
\centering
\includegraphics[width=\textwidth]{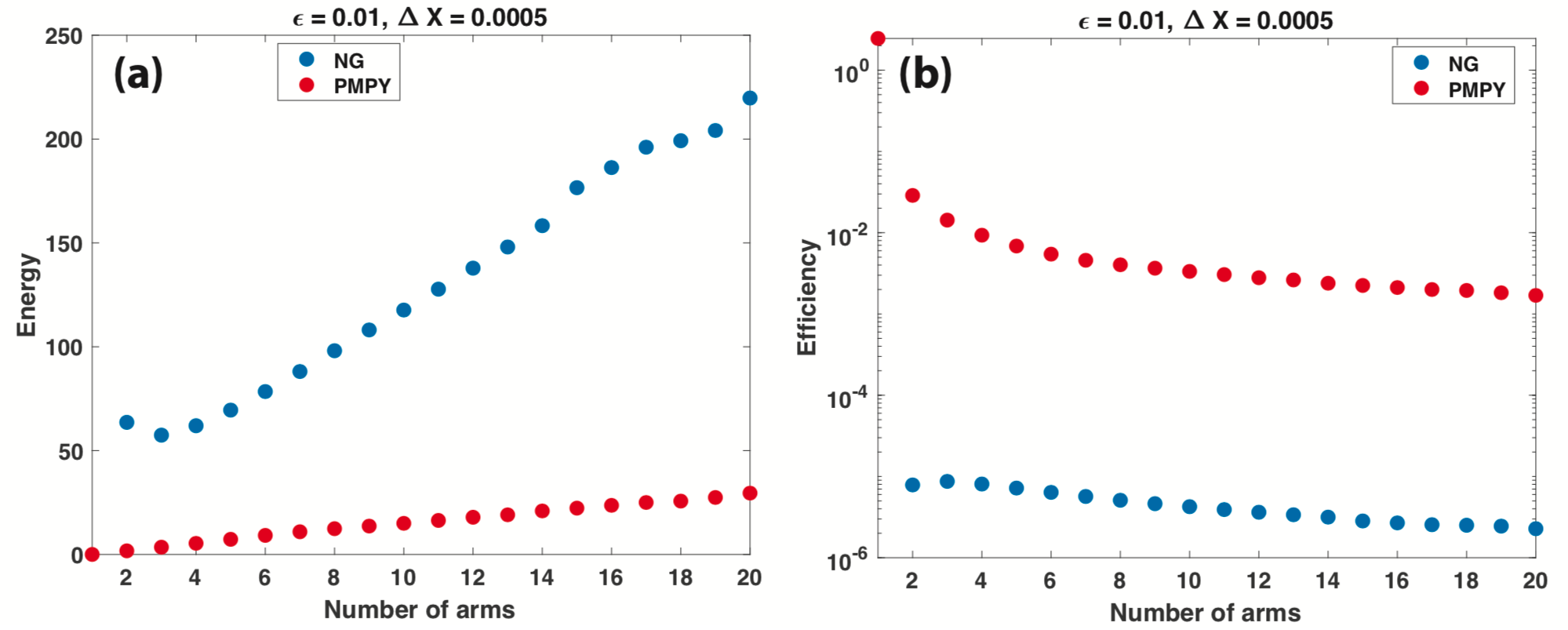}
\caption{\textbf{$(m+1)${-linked-sphere NG-type or PMPY-type of swimmers with widely separated spheres.}} (\textbf{a}) Energy dissipation of optimal strokes 
$\mathcal{E} (\overline{\gamma})$ and (\textbf{b}) efficiency Eff of NG (blue dots) and PMPY (red dots) swimmers with $m$ linking-arms and $m+1$ spheres. 
$\varepsilon =  {0.01}$ and $X = 5 \times 10^{-4}$ in all~simulations.
}
\label{Fig.NG-PMPY-NSph-L20}
\end{figure}

In this part we consider swimmers with widely separated spheres, where $\varepsilon = A/L = {0.01}$, comparing to $\varepsilon =0.2$
in Sections \ref{Sec.NG-NSph} and \ref{Sec.PMPY-NSph}. The simulation results are given in Figure \ref{Fig.NG-PMPY-NSph-L20} for NG and PMPY types of swimmers
with different numbers of linking-arms $m$, where $X = 5 \times 10^{-4}$ as before.

First, the most efficient NG type of swimmer is again the 4-sphere NG swimmer, with a minimum energy dissipation $\min_m \mathcal{E} (\overline{\gamma}) = {5745.09} \times 10^{-2}$; 
the most efficient PMPY type of swimmer is again the 2-sphere PMPY swimmer, with a minimum energy dissipation $\min_m \mathcal{E} (\overline{\gamma}) = {2.03} \times 10^{-2}$.
Comparing~to $\varepsilon = 0.2$ swimmers, we find that the optimized energy dissipation of NG swimmer with widely separated sphere increases greatly 
while that of PMPY swimmer is almost the same (Table \ref{Tab.Comp-NG-PMPY-NSph}).

\begin{table}[h]
\caption{Energy dissipation $\mathcal{E} (\overline{\gamma})$ of the most efficient NG and PMPY types of swimmers.}
\centering
\begin{tabular}{ccc}
\hline
	& \textbf{NG}	&   \textbf{PMPY} \\
	\hline
& \textbf{4-Sphere} \boldmath{($m=3$)} &  \textbf{2-Sphere} \boldmath{($m=1$)} \\
\hline
$\varepsilon = 0.2$ & $17.28 \times 10^{-2}$ & $1.70 \times 10^{-2}$ \\
$\varepsilon =  {0.01} $ & $ {5745.09 \times 10^{-2}}$ & ${2.03  \times 10^{-2}}$ \\
\hline
\end{tabular}
\label{Tab.Comp-NG-PMPY-NSph}
\end{table}

Next, similar to $\varepsilon = 0.2$ swimmers,  when $m$ increases, the energy dissipation $\mathcal{E} (\overline{\gamma})$ increases for both NG and PMPY types of swimmers (Figure \ref{Fig.NG-PMPY-NSph-L20}a), resulting in fast decreasing efficiency (Figure \ref{Fig.NG-PMPY-NSph-L20}b). However, we should point out that while for a 2-sphere PMPY swimmer, $\mathcal{E} (\overline{\gamma})$
is almost the same despite the separation distance of the spheres (Table \ref{Tab.Comp-NG-PMPY-NSph}), for large $m$, the energy dissipation of PMPY type of swimmers increases 
much faster with widely separated spheres ($\varepsilon = {0.01}$) comparing to those with closer spheres ($\varepsilon = 0.2$) (Figures \ref{Fig.N-link-NG}b and \ref{Fig.NG-PMPY-NSph-L20}a, Table \ref{Tab.Comp-NG-PMPY-21Sph}).

\begin{table}[h]
\caption{Energy dissipation $\mathcal{E} (\overline{\gamma})$ of 20-arms NG and PMPY types of swimmers.}
\centering
\begin{tabular}{ccc}
\hline
	& \textbf{NG 21-Spheres}	&   \textbf{PMPY 21-Spheres} \\
\hline
$\varepsilon = 0.2$ & $62.93  \times 10^{-2} $ & $50.29  \times 10^{-2} $ \\
$\varepsilon = {0.01}$ & $ { 21975.1 \times 10^{-2}}$ & $ {2954.27 \times 10^{-2}}$ \\
\hline
\end{tabular}
\label{Tab.Comp-NG-PMPY-21Sph}
\end{table}

\begin{remark}
{In~\cite{wang2015performance}, using asymptotic analysis, it is shown that for NG 3-sphere and VE 3-sphere swimmers, the
net translation $X (\gamma) \sim O (\varepsilon^2)$; while for PMPY 2-sphere swimmers, $X (\gamma) \sim O (1)$. Therefore as the 
separation distance $L$ increases, the efficiency of NG and VE 3-sphere swimmers quickly decreases, while that of PMPY 2-sphere
does not change much, which makes PMPY swimmers superior to NG and VE swimmers from an efficiency point of view, considering
that all have a minimal 2-degree of freedom in shape deformations. The~asymptotic analysis results supports our numerical results for optimal
strokes (Table \ref{Tab.Comp-NG-PMPY-NSph}). However, when~there are more degrees of freedom in a swimmer's shape deformations,
the efficiency behavior of PMPY types of swimmers clearly depends on the separation distance $L$ (Table \ref{Tab.Comp-NG-PMPY-21Sph}). 
Further asymptotic analysis in this direction might be worthwhile.
}
\end{remark}

\section{Discussion}

A successful and efficient locomotory gait design is important for micro-organisms and micro-robot who live or
present in a viscous fluid environment. While bacteria often adopt a flagellated or ciliated mode of swimming 
strategy, amoeboid cells deform their cell bodies to propel themselves, resisting the viscous resistance from 
surrounding fluid. Such shape deformations can be generalized into two modes: stretch or elongation of a part 
of or the whole cell body, and mass transportation along the the cell body which does not greatly change the 
cell length. Interestingly, in bio-engineering, there are three linked-sphere LRN swimming models (NG 3-sphere,
PMPY 2-sphere and VE 3-sphere models) that adopt a uni- or mixed modes of the aforementioned two shape 
deformation changes. By~analyzing the optimal strokes of the three swimmers, we show that PMPY which 
adopts the mixed control is the most efficient among the three. We also consider models consisting a chain 
of spheres, and again we find that the PMPY-type of swimmers that use mixed controls are more efficient than
NG-type of swimmers which use uni-controls of length change. We also find that generally speaking, the 
swimming efficiency decreases as the number of spheres increases, implying that more degrees of freedom
in shape deformations is not a good strategy in optimal gait design. When the sphere separation distance
increases, the efficiencies of NG type of swimmers greatly decrease, while the efficiencies of PMPY type of 
swimmers decrease in swimmers with many spheres, but are not affected much for swimmers with less spheres.

The findings in our paper can be potentially applied in the design of micro-robots with more complex structures. 
In addition, the numerical scheme presented here can be applied to more advanced LRN swimming systems. For example,
it can be applied for general 2D and 3D swimmers, when the swimmer shapes can be represented by 
conformal mappings or spherical harmonics~\cite{shapere1989geometry,wang2016computational,muskhelishvili2013some,
farutin2013amoeboid}.

Moreover, for micro-organisms, their swimming behaviors are also significantly affected by environmental factors in
biological media, therefore complex rheology of the surrounding fluid should be considered. A starting point might
be bringing the linked-sphere swimmers into viscoelastic medium. We would like to point out that asymptotic analysis
results for NG 3-sphere swimmer in linearized viscoelastic fluid have been obtained~\cite{curtis2013three}.

\vspace{6pt} 



\textbf{Funding: } This research received no external funding.
\textbf{acknowledgments.} The author is grateful to Dr. Hans Othmer for discussions on the research.


\textbf{Abbreviations:} The following abbreviations are used in this manuscript:\\

\noindent 
\begin{tabular}{@{}ll}
LRN & Low Reynolds Number\\
NG & Najafi-Golestanian 3-sphere model\\
PMPY & \textit{Pushmepullyou} 2-sphere model\\
VE & Volume-exchange 3-sphere model \\
SM & Shooting method
\end{tabular}

\appendix


\section{Shooting Method}\label{Sec.SM} 
We use the SM to solve the optimal stroke problem Equation (\ref{Eq.OptStr}), where the Euler--Lagrange Equation (\ref{Eq.EL})
is solved using the 2nd order Runge--Kutta method~\cite{alouges2008optimal, alouges2009optimal}.

For a LRN swimmer with finitely many degrees of freedom in its shape deformations, given~an initial shape $\gamma_0$ and
an initial shape deformation denoted by $v_0$, there exists a unique solution $\gamma (t) \  (t \in [0,1])$ to the Euler--Larange
Equation (\ref{Eq.EL}) and satisfies the initial condition: $\gamma (0) = \gamma_0$, $\dot{\gamma} (0) = v_0$. 
For~the given initial shape $\gamma_0$, define a function $\Phi_{\gamma_0}$ as:
\begin{eqnarray}\label{Eq.Phi}
\Phi_{\gamma_0} (v_0, \lambda) = \big( \gamma(1), X (\gamma) \big)
\end{eqnarray}
where $\lambda$ is the Lagrange multiplier in Equation (\ref{Eq.EL}), $\gamma$ is the solution to the Euler--Larange
Equation~(\ref{Eq.EL}), and $X (\gamma)$ is the net translation resulted from the 
stroke $\gamma$ (see Equation (\ref{Eq.NetTr-X})). Then the optimal stroke problem Equation (\ref{Eq.OptStr}) becomes:
\begin{quote}
Given initial shape $\gamma_0$ and a net translation $X_0$, find $(v_0, \lambda)$, s.t. $\Phi_{\gamma_0} (v_0, \lambda) = (\gamma_0, X_0)$.
\end{quote}

The corresponding stroke is the optimal stroke $\overline{\gamma}$.

\begin{remark}
{From the geometric point of view, let $Q$ be the set of all possible configurations of the LRN swimmer. If the swimmer has
only finitely many degrees of freedom in this shape deformations, then $Q$ is a finite dimensional smooth manifold and has 
the geometric structure $Q = S \times G$, where $S$ is the set of all unlocated shapes of the swimmer and $G$ is the group
of rigid motions. All the linked-sphere swimmers we considered are rotation-free and can only perform translation along one direction,
therefore $G \cong \mathbb{R}$. A swimming stroke $ \gamma (t)$ is a path in $S$, and $\dot{\gamma} (t) \in T_{\gamma (t)} S$,
that is, a shape deformation $\dot{\gamma} (t)$ is a tangent vector in the tangent space $T_{\gamma (t)} S$ on the shape $\gamma (t)$.
Therefore Equation (\ref{Eq.Phi}) defines a function $\Phi_{\gamma_0}$:
\begin{eqnarray*}
\Phi_{\gamma_0}: \ T_{\gamma_0} S \times \mathbb{R} \rightarrow Q
\end{eqnarray*}
and
\begin{eqnarray*}
v_0 \in T_{\gamma_0} S, \ \lambda \in \mathbb{R}, \ \gamma (1) \in S, \ X (\gamma) \in G \cong \mathbb{R}
\end{eqnarray*}
The geometric structure $Q = S \times G$ makes $Q$ the trivial principal fiber bundle over $S$. For more discussions regarding
the geometric properties of a LRN swimming system, please refer to 
\cite{shapere1989geometry, cherman2000low, loheac2013controllability, loheac2014controllability, chambrion2019optimal, avron2004optimal, leshansky2007frictionless}.}
\end{remark}

\begin{figure}[h]
\centering
\includegraphics[width=\textwidth]{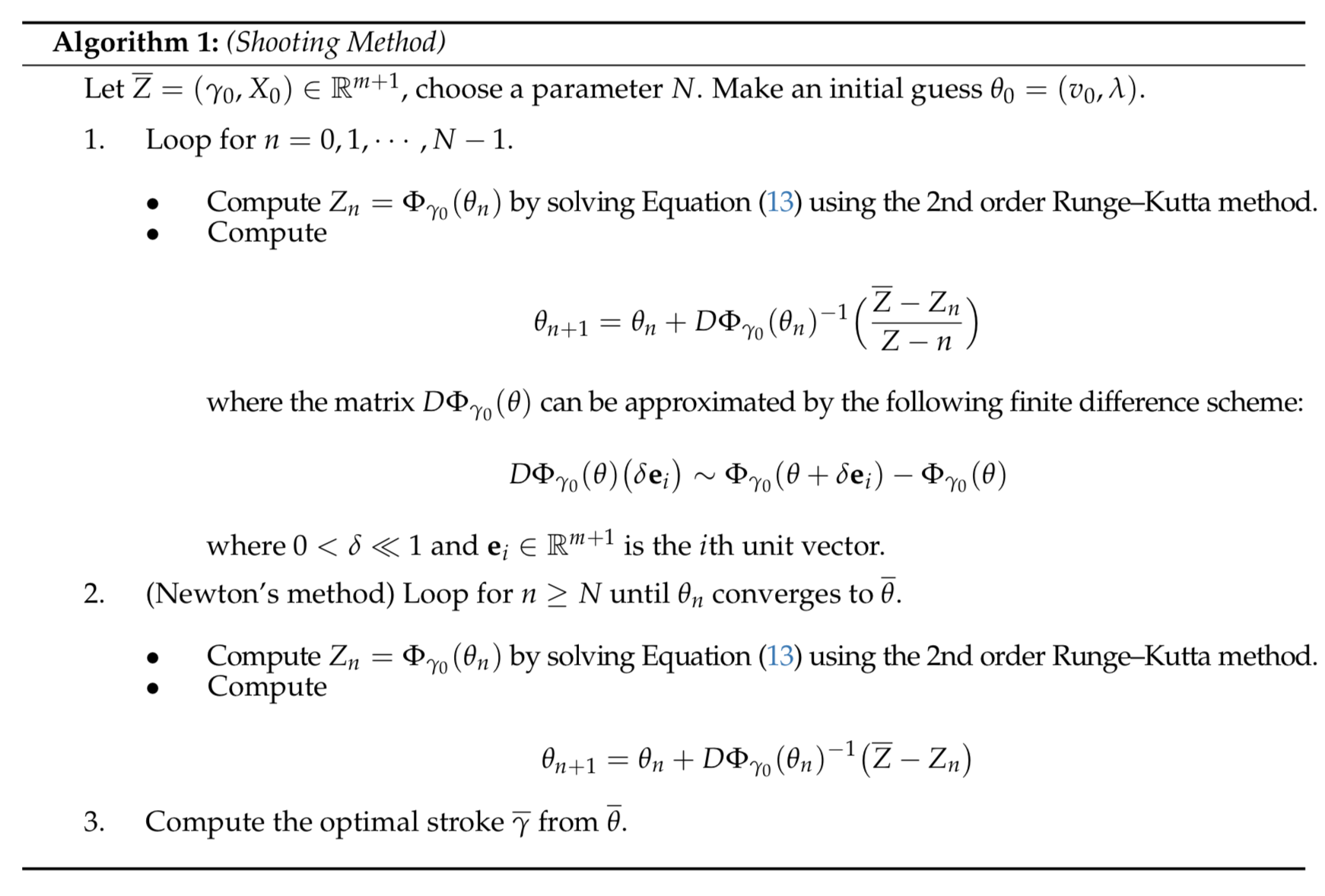}
\end{figure}

{
\section{Geodesic Strokes\label{Sec.GeoStroke}}

In~\cite{alouges2008optimal}, Alouges et al identified three types of geodesic strokes by their shapes, 
referred to as ``drop, bean and pretzel''. While ``drops'' and ``beans'' are single loop strokes, pretzel
can be considered as the general 2-loop strokes.

In our simulations of NG 3-sphere and PMPY 2-sphere models, we also capture the three types of 
geodesic strokes (Figure \ref{Fig.Geo-NG-PMPY-SS}, Tables \ref{Tab.Geo-NG-X045} and \ref{Tab.Geo-PMPY-X1}), 
where $X = 0.045, \ \varepsilon = 0.2$ for the NG swimmer and $X = 0.1, \  
\varepsilon = 0.2$ for the PMPY swimmer. In the NG swimmer, the ``drop" stroke gives the optimal stroke
among the three (Figure \ref{Fig.Geo-NG-PMPY-SS}a,b, Tables \ref{Tab.Geo-NG-X045}), and we note that in~\cite{alouges2008optimal} it is also reported that the ``drop'' stroke
is the most efficient one. On the other hand, in the PMPY swimmer, the ``drop'' and ``bean'' strokes 
consume similar energy (Figure \ref{Fig.Geo-NG-PMPY-SS}c,d, Tables \ref{Tab.Geo-PMPY-X1}). In either NG or PMPY model, single~loop geodesic strokes are clearly more 
efficient then the 2-loop geodesic strokes, or say, the~``pretzel'' strokes (Tables \ref{Tab.Geo-NG-X045} and \ref{Tab.Geo-PMPY-X1}).
Next, with large $X$, we catch multi loop geodesic strokes with more than 2 loops in the PMPY swimmer (Figure \ref{Fig.Geo-PMPY-X2-SS},
Table \ref{Tab.Geo-PMPY-X2}). Finally, We do not catch multi-loop geodesic strokes for VE swimmers, possibly
because that large shape deformations of VE models easily lead to negative volume of the spheres. 

\begin{figure}[h]
\centering
\includegraphics[width=\textwidth]{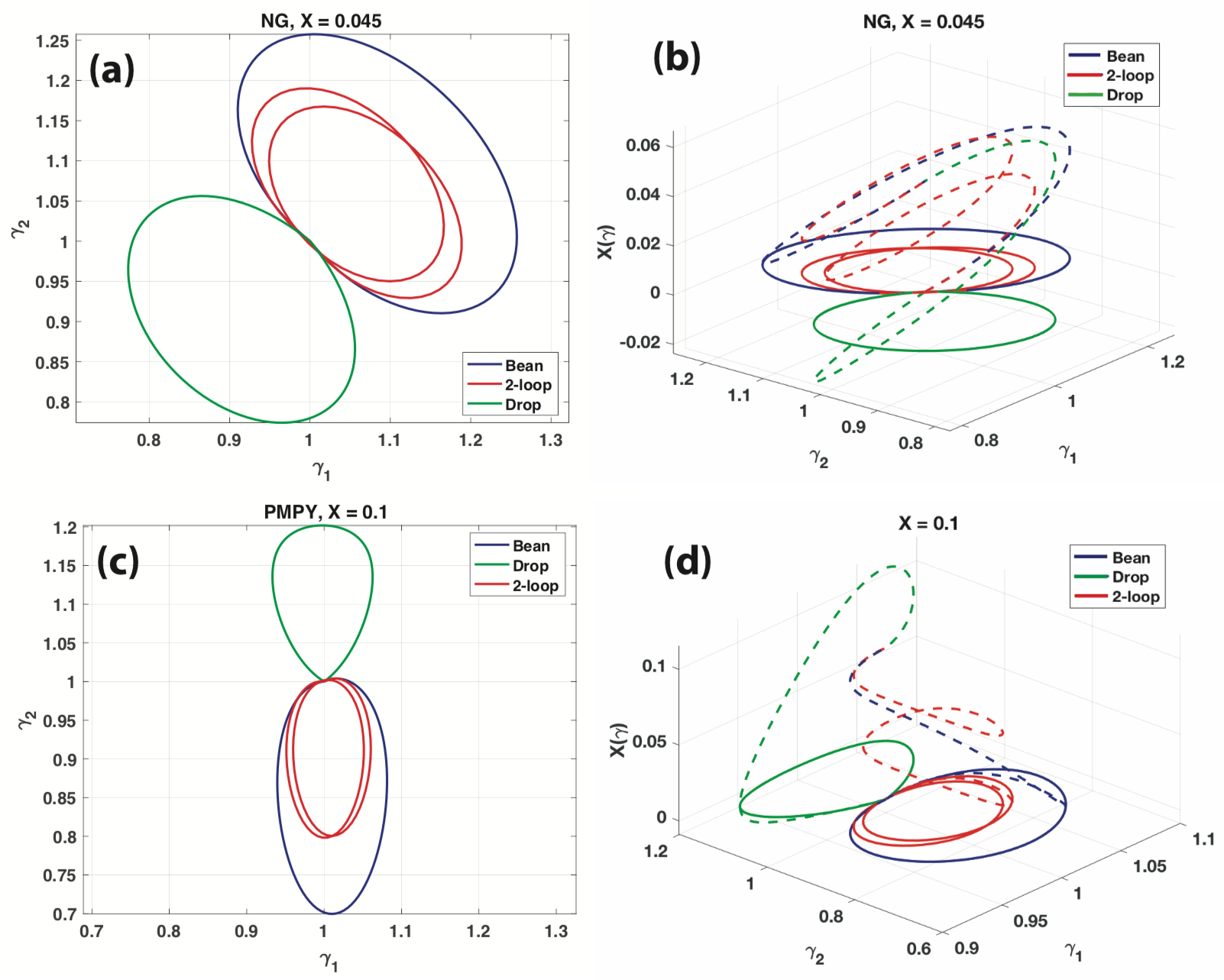}
\caption{\textbf{{Geodesic strokes of NG 3-sphere and PMPY 2-sphere swimmers.}} (\textbf{a}) The $\gamma$-paths of the geodesic strokes
of the NG 3-sphere swimmer, with $X(\gamma) = 0.045$ and $\varepsilon = 0.2$. \textbf{(b}) The geodesic stroke trajectories from (\textbf{a}).
(\textbf{c}) The $\gamma$-paths of the geodesic strokes
of the PMPY 2-sphere swimmer, with~$X(\gamma) = 0.1$ and $\varepsilon = 0.2$. (\textbf{d}) The geodesic stroke trajectories from (\textbf{c}).
}
\label{Fig.Geo-NG-PMPY-SS}
\end{figure}   

\begin{table}[h]
\caption{Energy dissipation $\mathcal{E} (\gamma)$ and efficiency Eff of geodesic strokes of the NG-3sphere swimmer,
with $X(\gamma) = 0.045, \ \varepsilon = 0.2$.}
\centering
\begin{tabular}{ccc}
\hline
\textbf{NG 3-Sphere}	& \boldmath{$\mathcal{E} (\gamma)$}	&   \textbf{Eff} \\
\hline
Bean & $1972.61  \times 10^{-2} $ & $2.28  \times 10^{-3} $ \\
Drop & $  1485.09 \times 10^{-2}$ & $3.03 \times 10^{-3}$ \\
2-loop (pretzel) & $3799.59 \times 10^{-2}$ & $ 1.18 \times 10^{-3}$ \\
\hline
\end{tabular}
\label{Tab.Geo-NG-X045}
\end{table}

\begin{table}[h]
\caption{Energy dissipation $\mathcal{E} (\gamma)$ and efficiency Eff of geodesic strokes of the PMPY 2-sphere swimmer,
with $X(\gamma) = 0.1, \ \varepsilon = 0.2$.}
\centering
\begin{tabular}{ccc}
\hline
\textbf{PMPY 2-Sphere}	& \boldmath{$\mathcal{E} (\gamma)$}	&  \textbf{ Eff} \\
\hline
Bean & $321.44  \times 10^{-2} $ & $31.12  \times 10^{-3} $ \\
Drop & $  320.94 \times 10^{-2}$ & $31.16 \times 10^{-3}$ \\
2-loop (pretzel) & $658.66 \times 10^{-2}$ & $ 15.19 \times 10^{-3}$ \\
\hline
\end{tabular}
\label{Tab.Geo-PMPY-X1}
\end{table}

\begin{figure}[h]
\centering
\includegraphics[width=\textwidth]{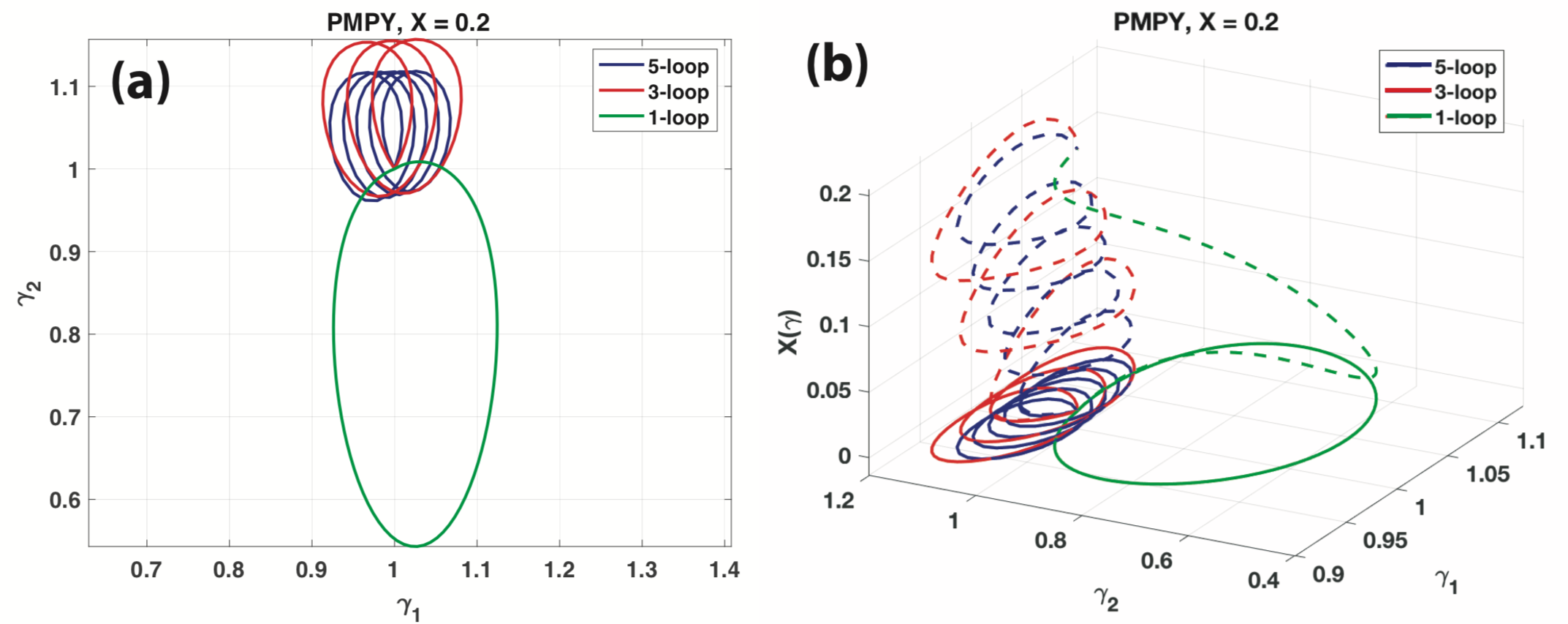}
\caption{\textbf{{Geodesic strokes of PMPY 2-sphere swimmers showing multi loops.}} (\textbf{a}) The $\gamma$-paths of the geodesic strokes
of the PMPY 2-sphere swimmer, with $X(\gamma) = 0.2$ and $\varepsilon = 0.2$. (\textbf{b}) The geodesic stroke trajectories from (\textbf{a}).
}
\label{Fig.Geo-PMPY-X2-SS}
\end{figure}   

\begin{table}[h]
\caption{Energy dissipation $\mathcal{E} (\gamma)$ and efficiency Eff of geodesic strokes of the PMPY 2-sphere swimmer,
with $X(\gamma) = 0.2, \ \varepsilon = 0.2$.}
\centering
\begin{tabular}{ccc}
\hline
\textbf{PMPY 2-Sphere}	& \boldmath{$\mathcal{E} (\gamma)$}	&   \textbf{Eff} \\
\hline
1-loop & $606.46  \times 10^{-2} $ & $32.99  \times 10^{-3} $ \\
3-loop & $  1917.31 \times 10^{-2}$ & $10.43 \times 10^{-3}$ \\
5-loop & $3197.44 \times 10^{-2}$ & $ 6.26 \times 10^{-3}$ \\
\hline
\end{tabular}
\label{Tab.Geo-PMPY-X2}
\end{table}

}

{
\section{Comparison between Optimal and Square Strokes\label{Sec.Comp-OptSq}}
Here we numerically compare the optimal stokes $\overline{\gamma}$ for NG 3-sphere, PMPY 2-sphere and VE 3-sphere models 
as obtained in Section \ref{Sec.CompNum_Org} with stepwise square strokes $\gamma_{\textrm{sq}}$. The energy dissipation
and efficiency are shown in Table \ref{Tab.Comp-Opt-SqLoop}, the $\gamma$-paths and the $X (\gamma (t)), t \in [0,1]$ trajectories are given in {Figure}~\ref{Fig.Comp-OptSq-SS}. 
To make a fair comparison, we take $\varepsilon = 0.2$ and $X (\gamma)  = 0.001$ in all simulations.
Data in Table~\ref{Tab.Comp-Opt-SqLoop} shows the improvement of swimming efficiency in the optimal strokes 
in all three model swimmers comparing to the stepwise square strokes.

\begin{table}[h]
\caption{Energy dissipation $\mathcal{E} (\gamma)$ and efficiency Eff of the optimal strokes $\overline{\gamma}$
and square strokes $\gamma_{\textrm{sq}}$. All simulations are taken with $X = 0.001$ and $\varepsilon = 0.2$.}
\centering
\begin{tabular}{ccccccc}
\hline
	& \multicolumn{2}{c}{\textbf{NG 3-Sphere}} 	&  \multicolumn{2}{c}{\textbf{PMPY 2-Sphere}} & \multicolumn{2}{c}{\textbf{VE 3-Sphere} } \\
	\hline
   & \boldmath{$\mathcal{E} $} & \textbf{Eff}   & \boldmath{$\mathcal{E}$} & \textbf{Eff}    & \boldmath{$\mathcal{E} $} & \textbf{Eff}    \\
\hline
Optimal strokes	& $37.61 \times 10^{-2} $  &  $ 2.66 \times 10^{-3} $ &$ 3.37 \times 10^{-2}$ & $29.71 \times 10^{-3}$	&$ 95.87 \times 10^{-2}$  & $1.04 \times 10^{-3}$ \\
($\overline{\gamma}$)            &  &      &  \\   
Square strokes	& $51.59 \times 10^{-2} $  &  $ 1.94 \times 10^{-3} $ &$ 5.04 \times 10^{-2}$ & $19.83 \times 10^{-3}$	&$ 181.93 \times 10^{-2}$  & $0.55 \times 10^{-3}$ \\
($\gamma_{\textrm{sq}}$)            &  &      &  \\   
\hline
\end{tabular}
\label{Tab.Comp-Opt-SqLoop}
\end{table}

\begin{figure}[h]
\centering
\includegraphics[width=\textwidth]{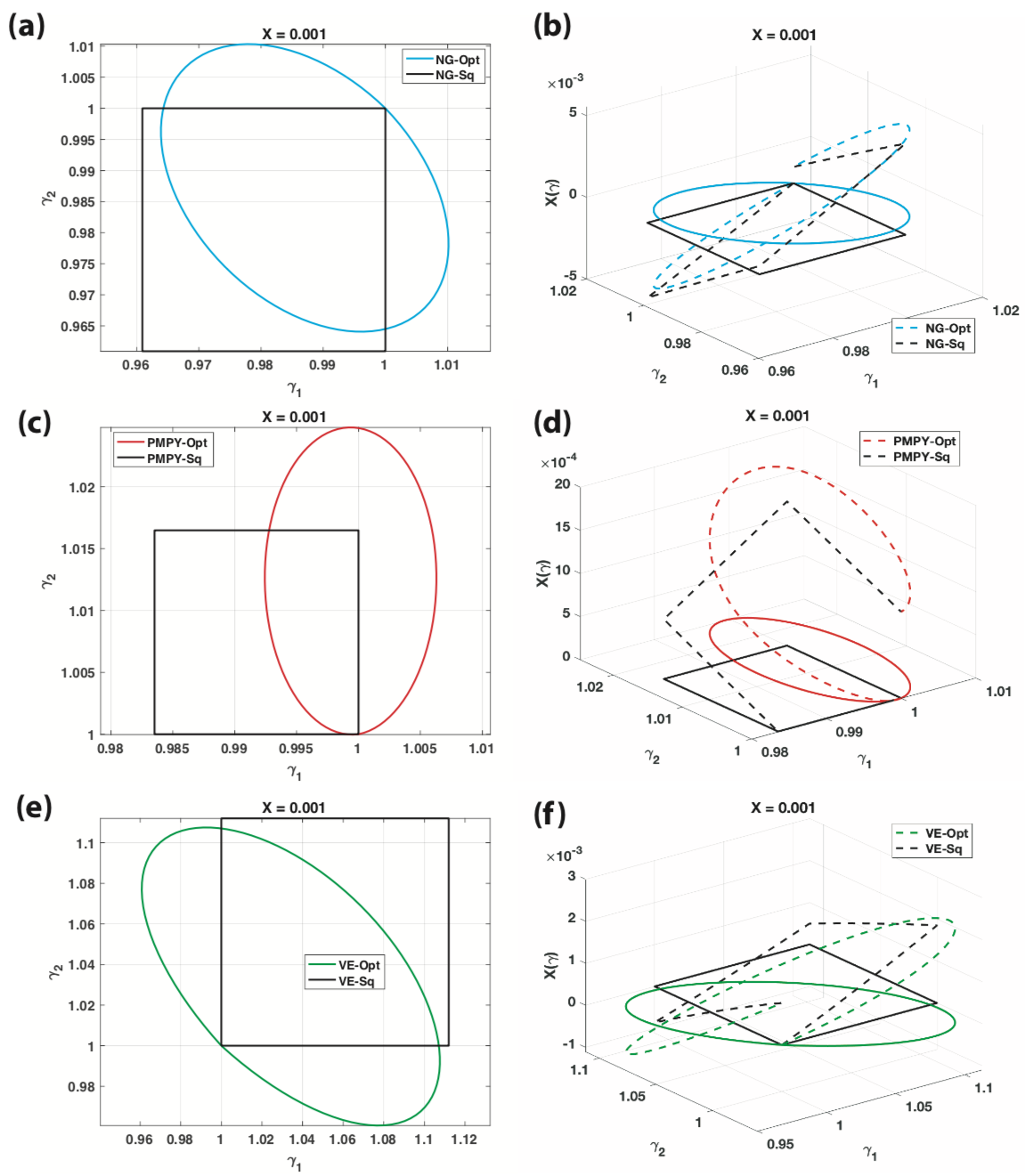}
\caption{\textbf{{Comparison between optimal and square strokes of NG 3-sphere, PMPY 2-sphere and VE 3-sphere swimmers.}} (\textbf{a},\textbf{c},\textbf{e}) 
give the $\gamma$-paths of the optimal strokes ($\overline{\gamma}$) and the square strokes ($\gamma_{\textrm{sq}}$)
of the NG, PMPY and VE swimmers, all with $X(\gamma) = 0.001 $ and $\varepsilon = 0.2$.
(\textbf{b},\textbf{d},\textbf{f}) give the stroke trajectories $X( \overline{\gamma} (t))$, $t \in [0,1]$ (dashed lines) corresponding to the $\gamma$-paths (solid lines).
}
\label{Fig.Comp-OptSq-SS}
\end{figure}

}




\clearpage
\begin{thebibliography}{999}

\bibitem{purcell1977life}
Purcell, E.M. Life at low Reynolds number. {\em Am. J. Phys.} {\bf 1977}, {\em 45}, 3--11.

\bibitem{lauga2006swimming}
Lauga, E.; Willow, R.; DiLuzio, G.M. Whitesides, and Howard A. Stone. Swimming in circles: Motion of bacteria near solid boundaries. 
{\em Biophys. J.} {\bf 2006}, {\em 90}, 400--412.

\bibitem{rafai2010effective}
Rafa{\"i}, S.; Levan J.; Philippe, P. 
{Effective viscosity of microswimmer suspensions. 
{\em Phys. Rev. Lett.} {\bf 2010}, {\em 104},~098102.}

\bibitem{riedel2005self}
Riedel, I.H.; Karsten, K.; Jonathon, H. 
A self-organized vortex array of hydrodynamically entrained sperm cells. 
{\em Science} {\bf 2005}, {\em 309}, 300--303.

\bibitem{sokolov2010swimming}
Sokolov, A.; Apodaca, M.M.; Grzybowski, B.A.; Aranson, I.S. 
Swimming bacteria power microscopic gears. 
{\em PNAS} {\bf 2010}, {\em 107}, 969--974.

\bibitem{van2011amoeboid}
Van Haastert, P.J. 
{Amoeboid cells use protrusions for walking, gliding and swimming.
{\em PLoS ONE} {\bf 2011}, {\em 6}, e27532.}

\bibitem{barry2010dictyostelium}
Barry, N.P.; Mark, S.B. 
Dictyostelium amoebae and neutrophils can swim. 
{\em PNAS} {\bf 2010}, {\em 107}, 11376--11380.

\bibitem{franz2018fat}
Franz, A.; Wood, W.; Martin, P. Fat body cells are motile and actively migrate to wounds to drive repair and prevent infection. 
{\em Dev. Cell.} {\bf 2018}, {\em 44}, 460--470.

\bibitem{nelson2010microrobots}
Nelson, B.J.; Kaliakatsos, I.K.; Abbott, J.J. Microrobots for minimally invasive medicine.
{\em Annu. Rev. Biomed. Eng.} {\bf 2010}, {\em 12}, 55--85.

\bibitem{najafi2004simple}
Najafi, A.; Golestanian, R. 
Simple swimmer at low Reynolds number: Three linked spheres. 
{\em Phys. Rev. E} {\bf 2004}, {\em 69}, 062901.




\bibitem{alexander2009hydrodynamics}
Alexander, G.P.; Pooley, C.M.; Yeomans, J.M. 
Hydrodynamics of linked sphere model swimmers. 
{\em J.~Phys.-Condens. Matter} {\bf 2008}, {\em 21}, 204108.

\bibitem{golestanian2008analytic}
Golestanian, R.; Ajdari, A. 
Analytic results for the three-sphere swimmer at low Reynolds number. 
{\em Phys.~Rev.~E} {\bf 2008}, {\em 77}, 036308.

\bibitem{curtis2013three}
Curtis, M.P.; Gaffney, E.A. 
Three-sphere swimmer in a nonlinear viscoelastic medium. 
{\em Phy. Rev. E} {\bf 2013},~{\em 87},~043006.

\bibitem{alouges2018parking}
Alouges, F.; Di Fratta, G.
Parking 3-sphere swimmer I. Energy minimizing strokes.
{\em Discre. Cont. Dyn-B} {\bf 2018},~{\em 23}, 1797--1817.

\bibitem{avron2005pushmepullyou}
{Avron, J.E.; Kenneth, O.; Oaknin, D.H. 
Pushmepullyou: An efficient micro-swimmer. 
{\em New J. Phys.} {\bf 2005}, {\em 7}, 234.}

\bibitem{rizvi2018three}
Rizvi, M.S.; Farutin, A.; Misbah, C. 
Three-bead steering microswimmers. 
{\em Phys. Rev. E} {\bf 2018}, {\em 97}, 023102.

\bibitem{wang2012models}
Wang, Q.; Hu, J.; Othmer, H. 
\emph{Natural Locomotion in Fluids and on Surfaces: Swimming, Flying, and Sliding};
chapter Models of low reynolds number swimmers inspired by cell blebbing; 
Springer:{Berlin, Germany}, 2012; pp. 185--195.

\bibitem{zhang2010experimental}
Zhang, S.; Or, Y.; Murray, R.M. 
Experimental demonstration of the dynamics and stability of a low Reynolds number swimmer near a plane wall.
In Proceedings of the 2010 American Control Conference, {Baltimore,~MD, USA, 30 June--2 July 2010}; pp. 1013--1041.


\bibitem{or2011dynamics}
Or, Y.; Zhang, S.; Murray, R.M. 
Dynamics and stability of low-Reynolds-number swimming near a wall. 
{\em Siam~J. Appl. Dyn. Syst.} {\bf 2011}, {\em 10}, 1013--1041.

\bibitem{saadat2019experimental}
Saadat, M.; Mirzakhanloo, M.; Shen, J.; Tomizuka, M.; Alam, M.R. The Experimental Realization of an Artificial Low-Reynolds-Number Swimmer with Three-Dimensional Maneuverability. 
{\em arXiv} {\bf 2019}, arXiv:1905.05893.



\bibitem{kadam2017trajectory}
Kadam, S.; Joshi, K.; Gupta, N.; Katdare, P.; Banavar, R. 
Trajectory tracking using motion primitives for the purcell's swimmer. 
In Proceedings of the 2017 IEEE/RSJ International Conference on Intelligent Robots and Systems (IROS), {Vancouver, BC, Canada, 24--28 September 2017}; pp. 3246--3251.

\bibitem{alouges2008optimal}
Alouges, F.; Antonio, D.; Aline, L. 
Optimal strokes for low Reynolds number swimmers: An example. 
{\em J.~Nonlinear~Sci.} {\bf 2008}, {\em 18}, 277--302.

\bibitem{alouges2009optimal}
Alouges, F.; Antonio, D.; Lefebvre, A.
Optimal strokes for axisymmetric microswimmers.
{\em Eur. Phys. J. E} {\bf 2009}, {\em 28}, 279--284.

\bibitem{alouges2011numerical}
Alouges, F.; DeSimone, A.; Heltai, L.
Numerical strategies for stroke optimization of axisymmetric microswimmers.
{\em Math. Mod. Meth. Appl. S} {\bf 2011}, {\em 21}, 361--387.

\bibitem{avron2004optimal}
{Avron, J.E.; Gat, O.; Kenneth, O. 
Optimal swimming at low Reynolds numbers.
{\em Phys. Rev. Lett.} {\bf 2004}, {\em 93}, 186001.}







\bibitem{avron2008geometric}
Avron, J.E.; Raz, O. 
A geometric theory of swimming: Purcell's swimmer and its symmetrized cousin. 
{\em New~J.~Phys.} {\bf 2008}, {\em 10}, 063016.

\bibitem{blake1971self}
Blake, J.R. 
Self propulsion due to oscillations on the surface of a cylinder at low Reynolds number. 
{\em B Aust. Math. Soc.} {\bf 1971}, {\em 5}, 255--264.

\bibitem{chambrion2019optimal}
Chambrion, T.; Giraldi, L.; Munnier, A.
Optimal strokes for driftless swimmers: A general geometric approach. 
{\em ESAIM Control Optim. Calc.} {\bf 2019}, \emph{25}, 6.

\bibitem{ishimoto2014swimming}
Ishimoto, K.; Gaffney, E.A. 
Swimming efficiency of spherical squirmers: beyond the Lighthill theory. 
{\em Phys.~Rev.~E} {\bf 2014}, {\em 90}, 012704.

\bibitem{lighthill1952squirming}
Lighthill, M.J. On the squirming motion of nearly spherical deformable bodies through liquids at very small Reynolds numbers. 
{\em Commun. Pur. Appl. Math.} {\bf 1952}, {\em 5}, 109--118.


\bibitem{loheac2013controllability}
Loh{\'e}ac, J.; Scheid, J.-F.; Tucsnak, M. 
Controllability and time optimal control for low Reynolds numbers swimmers. 
{\em Acta Appl. Math.} {\bf 2013}, {\em 123}, 175--200.

\bibitem{loheac2014controllability}
Loh{\'e}ac, J.; Munnier, A. 
Controllability of 3D low Reynolds number swimmers 
{\em ESAIM Control Optim. Cacl.} {\bf 2014}, {\em 20}, 236--268.

\bibitem{shapere1989efficiencies}
Shapere, A.; Wilczek, F. 
Efficiencies of self-propulsion at low Reynolds number. 
{\em J. Fluid. Mech.} {\bf 1989}, {\em 198}, 587--599.

\bibitem{tsang2018self}
Tsang, A.C.H.; Tong, P.W.; Nallan, S.; Pak, O.S.
Self-learning how to swim at low Reynolds number.
{\em arXiv}~{\bf 2019}, arXiv:1808.07639.



\bibitem{shapere1989geometry}
Shapere, A.; Wilczek, F. Geometry of self-propulsion at low Reynolds number. {\em J. Fluid. Mech.} {\bf 1989}, {\em 198}, 557--585.

\bibitem{pozrikidis1992boundary}
Pozrikidis, C. {\em Boundary Integral and Singularity Methods for Linearized Viscous Flow}; Cambridge University Press: {Cambridge, UK}, 1992.

\bibitem{kim2013microhydrodynamics}
Kim, S.; Karrila, S.J. {\em Microhydrodynamics: Principles and Selected Applications}; Courier Corporation: {North Chelmsford, MA, USA}, 2013.

\bibitem{liang2013fast}
Liang, Z.; Gimbutas, Z.; Greengard, L.; Huang, J.; Jiang, S. 
A fast multipole method for the Rotne--Prager--Yamakawa tensor and its applications. 
{\em J. Comput. Phys. Mech.} {\bf 2014}, {\em 234}, 133--139.

\bibitem{wajnryb2013generalization}
Wajnryb, E.; Mizerski, K.A.; Zuk, P.J.; Szymczak, P. Generalization of the Rotne--Prager--Yamakawa
mobility and shear disturbance tensors. {\em J. Fluid. Mech.} {\bf 2013}, {\em {731}}.

\bibitem{wang2018analysis}
Wang, Q.; Othmer, H.G. 
Analysis of a model microswimmer with applications to blebbing cells and mini-robots. 
{\em J. Math. Biol.} {\bf 2018}, {\em 76}, 1699--1763.

\bibitem{yamakawa1970transport}
Yamakawa, H. Transport properties of polymer chains in dilute solution: hydrodynamic interaction. {\em J. Chem. Phys.} {\bf 1970}, {\em 53}, 436--443.

\bibitem{zuk2014rotne}
Zuk, P.J.; Wajnryb, E.; Mizerski, K.A.; Szymczak, P. Rotne--Prager--Yamakawa approximation for different-sized particles in application to macromolecular bead models. 
{\em J. Fluid. Mech.} {\bf 2014}, {\em 741}, 436--443.

\bibitem{cherman2000low}
Cherman, A.; Delgado, J.; Duda, F.; Ehlers, K.; Koiller, J.; Montgomery, R. 
Low Reynolds number swimming in two dimensions. 
In {\em Hamiltonian Systems And Celestial Mechanics: (HAMSYS--98)};
World Scientific: {Singapore}, 2000; pp. 32--62.

\bibitem{leshansky2007frictionless}
Leshansky, A.M.; Kenneth, O.; Gat, O.; Avron, J.E. 
A frictionless microswimmer. 
{\em New J. Phys.} {\bf 2007}, {\em 9}, 145.

\bibitem{nasouri2019efficiency}
Nasouri, B.; Vilfan, A.; Golestanian, R.
Efficiency limits of the three-sphere swimmer.
{\em arXiv} {\bf 2019}, arXiv:1905.06510.

\bibitem{elgeti2015physics}
Elgeti, J.; Winkler, R.G.; Gompper, G. Physics of microswimmers---Single particle motion and collective behavior: A review. 
{\em Rep. Prog. Phys.} {\bf 2015}, {\em 78}, 056601.

\bibitem{hancock1953self}
Hancock, G.J. The self-propulsion of microscopic organisms through liquids. {\em Proc. R. Soc. Lond. A} {\bf 1953}, {\em 217}, 96--121.

\bibitem{higdon1979hydrodynamics}
Higdon, J.J.L. The hydrodynamics of flagellar propulsion: helical waves. {\em J. Fluid Mech.} {\bf 1979}, {\em 94}, 331--351.

\bibitem{lauga2009hydrodynamics}
Lauga, E.; Powers, T.R. The hydrodynamics of swimming microorganisms. 
{\em Rep. Prog. Phys.} {\bf 2009}, {\em 72}, 096601.

\bibitem{patteson2015running}
Patteson, A.E.; Gopinath, A.; Goulian, M.; Arratia, P.E. Running and tumbling with E. coli in polymeric solutions. 
{\em Sci. Rep.} {\bf 2015}, {\em 5}, 15761.

\bibitem{phan1987boundary}
Phan-Thien, N.; Tran-Cong, T.; Ramia, M. A boundary-element analysis of flagellar propulsion. {\em J. Fluid Mech.} {\bf 1987}, {\em 184}, 533--549.

\bibitem{sowa2008bacterial}
Sowa, Y.; Richard, M.B. Bacterial flagellar motor. 
{\em Q. Rev. Biophys.} {\bf 2008}, {\em 41}, 103--132.

\bibitem{taylor1952action}
Taylor, G.I. The action of waving cylindrical tails in propelling microscopic organisms. {\em Proc. R. Soc. Lond.} {\bf 1952}, {\em 211}, 225--239.

\bibitem{bae2010swimming}
Bae, A.J.; Eberhard, B. On the swimming of Dictyostelium amoebae. 
{\em PNAS} {\bf 2010}, {\em 107}, E165--E166.

\bibitem{wang2015performance}
Wang, Q.; Othmer, H.G. 
The performance of discrete models of low reynolds number swimmers. 
{\em Math.~Biosci.~Eng.} {\bf 2015}, {\em 12}, 1303--1320.

\bibitem{farutin2013amoeboid}
Farutin, A.; Rafa{\"\i}, S.; Dysthe, D.K.; Duperray, A.; Peyla, P.; Misbah, C. 
Amoeboid swimming: A generic self-propulsion of cells in fluids by means of membrane deformations. 
{\em Phy. Rev. L} {\bf 2013}, {\em 111}, 228102.

\bibitem{muskhelishvili2013some}
{Muskhelishvili, N.I. 
{\em Some Basic Problems of the Mathematical Theory of Elasticity}; Springer: {Berlin, Germany}, 2013.}

\bibitem{wang2016computational}
Wang, Q.; Othmer, H.G. 
Computational analysis of amoeboid swimming at low Reynolds number. 
{\em J.~Math.~Biol.} {\bf 2016}, {\em 72}, 1893--1926.



\end{thebibliography}
\end{document}